%% file: main.tex
\documentclass[letterpaper]{article} 
\usepackage{aaai23}  
\usepackage{times}  
\usepackage{helvet}  
\usepackage{courier}  
\usepackage[hyphens]{url}  
\usepackage{graphicx} 
\usepackage{natbib}  
\usepackage{caption} 
\usepackage{algorithm}
\usepackage{algorithmic}

\usepackage{float}
\usepackage{amsmath}
\usepackage{xcolor}         

\usepackage{newfloat}
\usepackage{listings}
\DeclareCaptionStyle{ruled}{labelfont=normalfont,labelsep=colon,strut=off} 
\floatstyle{ruled}
\newfloat{listing}{tb}{lst}{}
\floatname{listing}{Listing}
\title{Where Does My Model Underperform?\\ 
A Human Evaluation of Slice Discovery Algorithms}
\author{
    Nari Johnson\textsuperscript{\rm 1},
    Ángel Alexander Cabrera\textsuperscript{\rm 1},
    Gregory Plumb\textsuperscript{\rm 1,2}\footnote{The author completed the majority of their work while a student at CMU.},
    Ameet Talwalkar\textsuperscript{\rm 1}
}
\affiliations{
    \textsuperscript{\rm 1}Carnegie Mellon University\\
    \textsuperscript{\rm 2}Inpleo, Inc.\\

    narij@andrew.cmu.edu, cabrera@cmu.edu, gdplumb@alumni.cmu.edu, talwalkar@cmu.edu
}

\usepackage{bibentry}

\newcommand{\ie}{\textit{i.e.,}}
\newcommand{\eg}{\textit{e.g.,}}

\begin{document}

\maketitle

\begin{figure*}[h]
\centering
\includegraphics[width=\linewidth]{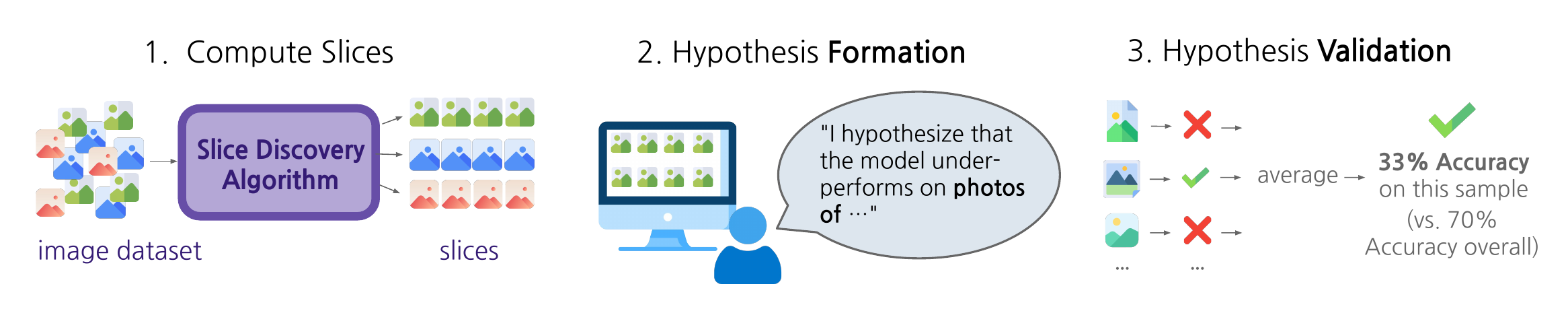}
\caption{An overview of our user study in three steps.  (Left) We run different slice discovery algorithms to compute high-error slices (subsets of an input dataset).  (Middle) We conduct a human subject study where we \emph{show the slices from Step 1} to a human subject, who forms a hypothesis (\ie{} description of a subgroup where the model underperforms) corresponding to each slice.  (Right) We validate each user-generated hypothesis from Step 2 by calculating the model's accuracy on a new sample of images that match the user's description.}
\label{fig:experiment-phases}
\end{figure*}

\begin{abstract}
Machine learning (ML) models that achieve high average accuracy can still underperform on semantically coherent subsets (``slices'') of data.
This behavior can have significant societal consequences for the safety or bias of the model in deployment, but identifying these underperforming slices can be difficult in practice, especially in domains where practitioners lack access to group annotations to define coherent subsets of their data.
Motivated by these challenges, ML researchers have developed new \emph{slice discovery algorithms} that aim to group together coherent and high-error subsets of data.
However, there has been little evaluation focused on whether these tools help humans form correct hypotheses about where (for which groups) their model underperforms.
We conduct a controlled user study ($N = 15$) where we show $40$ slices output by two state-of-the-art slice discovery algorithms to users, and ask them to form hypotheses about an object detection model.
Our results provide positive evidence that these tools provide some benefit over a naive baseline, and also shed light on challenges faced by users during the hypothesis formation step.
We conclude by discussing design opportunities for ML and HCI researchers.
Our findings point to the importance of centering users when creating and evaluating new tools for slice discovery.
\end{abstract}

\input{text/hcomp-sections/introduction}

\input{text/hcomp-sections/slice-overview}

\input{text/hcomp-sections/related-work}

\input{text/hcomp-sections/methodology}

\input{text/hcomp-sections/results}

\input{text/hcomp-sections/discussion}

\input{text/hcomp-sections/conclusion}

\input{text/hcomp-sections/acknowledgements}

\bibliography{aaai23}

\newpage
\appendix
\input{text/hcomp-sections/appendix}

\end{document}

%% file: text/hcomp-sections/introduction.tex
\section{Introduction}

A growing number of works propose tools to help stakeholders form hypotheses about the behavior of machine learning (ML) models.
One type of behavior that can have significant societal consequences occurs when a model underperforms on semantically coherent subsets (\ie{} ``slices'') of data.
For example, \citet{buolamwini2018gender} found that leading tech companies' commercial facial recognition models were significantly less accurate at classifying faces of women with darker skin.
Knowledge of this model behavior informed advocacy efforts that led to fundamental changes to dataset curation for facial recognition \citep{ birhane2022unseen} and public deliberation surrounding governance of facial recognition systems \citep{raji2022actionable}.
More broadly, knowledge of underperforming slices can inform model selection and deployment \citep{balayn2023faulty} or actions that can be taken to fix the model \citep{holstein2019improving, cabrera2021discovering, idrissi2022simple}.

However, identifying these underperforming slices can be difficult in practice. 
In many domains, practitioners often do not have access to group annotations that can be used to define semantically \emph{coherent} (\ie{} united by a single human-understandable concept) subsets of their data \citep{cabrera2022what}.
Motivated by these challenges, ML researchers have developed new automated tools in the growing field of \emph{slice discovery}.
At a high level, these \emph{slice discovery algorithms} are unsupervised methods that aim to group together coherent and high-error slices of data \citep{ sohoni2020no, d2021spotlight, singla2021understanding, eyuboglu2022domino, plumb2023rigorous, wang2023error}.
These works propose that a human stakeholder can inspect the slices output by these algorithms to form hypotheses of model behavior, \ie{} describe in words a group where the model underperforms.

While researchers continue to develop new slice discovery algorithms, there has been little evaluation of whether these algorithms help stakeholders achieve their proposed goals.
Past human evaluations of slice discovery tools have used subjective judgments of the output slices' coherence (such as whether users can find a description that matches the majority of images in the slice) as proxies of these algorithms' utility \citep{singla2021understanding, d2021spotlight}, but stop short of verifying whether these descriptions are useful or even accurate depictions of the model's behavior.
In this work, we ask: Do the slices output by these algorithms help users form \emph{correct hypotheses of model behavior?}

As a motivating example, consider a scenario where a practitioner wishes to evaluate a new object detection model designed to be deployed in an autonomous vehicle.
The practitioner could run a slice discovery algorithm on a dataset of dash-cam photos collected from field testing, but its output (groups of photos) is not immediately actionable.
Describing in words a group where the model underperforms (\eg{} ``the model fails to identify stop signs in snowy environments'') is a prerequisite step for several downstream actions that the practitioner's company could take, such as initiating targeted data collection efforts (\eg{} additional field testing to gather more data in snowy regions) or selective deployment (\eg{} delaying roll-out of the new model in areas that experience snow).

Unfortunately, taking action to address an \emph{incorrect} hypothesis can have several undesirable consequences.
For example, if the model actually performs just as well in snowy environments, then the company's efforts to improve the model's performance on that group may have been better expended elsewhere, or may even worsen the model's performance on other groups \citep{li2022whac}.
Thus, we argue that this under-examined step of hypothesis formation is critically important for many stakeholders.

In this work, we design a controlled user study to investigate how participants make sense of the slices output by algorithms to form hypotheses of model behavior.
Our study has three parts, illustrated in Figure \ref{fig:experiment-phases}.
First, after training a model on a large object detection dataset \citep{lin2014microsoft}, we run two state-of-the-art slice discovery algorithms, \textsc{Domino} \citep{eyuboglu2022domino} and \textsc{PlaneSpot} \citep{plumb2023rigorous} that output high-error slices of data.
Next, we show these slices to $N = 15$ study participants and ask them to \emph{form behavioral hypotheses} (\ie{} describe in words a group where the model underperforms) corresponding to each slice.
Finally, we then \emph{validate each hypothesis} (\ie{} whether the model actually underperforms on these groups) by measuring the model's performance on new data that matches each hypothesis.  
In summary, our work makes the following contributions:

\begin{itemize}
    \item \emph{Benchmarking existing slice discovery tools}. Our study results provide positive evidence that existing tools can sometimes (though not always) help stakeholders form correct hypotheses about where their model underperforms.
    Specifically, participants were more likely to form correct hypotheses when shown slices output by \textsc{Domino} and \textsc{PlaneSpot}, relative to a naive baseline condition where they were shown a random sample of misclassified images.
    \item \emph{Characterizing how users may (mis)interpret existing tools}.  Our analyses shed light onto how the slices output by these tools may be misleading.
    First, we found \emph{no significant association} between the number of images in a slice that the user selected as ``matching'' their hypothesis, and its correctness.
    This result challenges conventional wisdom that slices that are coherent (\ie{} the majority of images in the slice match a common description) do in fact correspond to a true model error (\ie{} the model underperforms on all such images).
    Second, we found that different users often formed different hypotheses when shown the same slice.
    Taken together, these findings illuminate the nuance and challenges faced by users during the hypothesis formation step.

    \item \emph{Design opportunities for future tools}.  Our findings point to several exciting design opportunities for ML and HCI researchers.  
    We highlight open challenges and possible paths forward to better support users as they make sense of where their model underperforms.
\end{itemize}


%% file: text/hcomp-sections/slice-overview.tex
\section{Slice Discovery Preliminaries}

In this section, we present an overview of the slice discovery problem, formalize the hypothesis formation step, and discuss our approach to validate each hypothesis.

\subsection{2.1 {}{} Slice Discovery Algorithms}\label{sec:sdms}

Many tools have been proposed to help users discover where their model underperforms, including post-hoc explanations \citep{kim2018interpretability, adebayo2022post}, methods that generate counterfactual data \citep{wiles2023discovering}, and data visualization interfaces \citep{cabrera2023zeno, suresh2023kaleidoscope, moore2023failure}.
In our work, we focus on \emph{slice discovery algorithms}: automated methods that aim to partition the data into coherent and high-error subsets \citep{ eyuboglu2022domino, plumb2023rigorous}.  
We exclude methods that rely on additional information such as fine-grained group annotations \citep{polyzotis2019slice, liu2021just} or human supervision to define coherent subsets.
Consequently, the methods we study do \emph{not} require the practitioner to anticipate the types of inputs where the model may underperform in advance.

\paragraph{Definition} We follow \citet{plumb2023rigorous} and define a slice discovery algorithm as a method that given as input a trained model $f$ and dataset of labeled images $D = \{(x_i, y_i) \}_{i = 1}^n$\footnote{We follow past work and give a separate held-out test set that the model $f$ was \emph{not} trained on as the input to a slice discovery algorithm.}, outputs $k$ slices $\left[ \Psi_j \right]_{j = 1}^k$.  
Each slice $\Psi_j \subseteq D$ is a subset of the input data.
The objective of slice discovery is for each slice $\Psi_j$ to correspond to a single coherent group where the model indeed underperforms (\ie{} $f$ has lower accuracy for images in $\Psi_j$ than for all images in $D$).
Past works propose that a user can inspect the datapoints belonging to each slice to ``\emph{identify common attributes}'' \citep{eyuboglu2022domino} to describe the underperforming groups.

\paragraph{Methods}
We evaluate two slice discovery algorithms: \textsc{Domino} \citep{eyuboglu2022domino} and \textsc{PlaneSpot} \citep{plumb2023rigorous}.  
Both algorithms work by running error-aware clustering on an embedding of each image, and differ in the specific clustering algorithm and embedding that they use.
We describe each algorithm in detail in Appendix A.
We evaluate these two algorithms because they achieved state-of-the-art performance on past benchmarks  \citep{eyuboglu2022domino, plumb2023rigorous} at the time we ran our study.

Our \textsc{Baseline} algorithm randomly samples from the set of misclassified images, without replacement.
While a subset of random misclassified images is high-error, it is unlikely to be coherent.
Our baseline was designed to simulate a naive workflow where a practitioner inspects an unsorted sample of misclassified images to form hypotheses.

\subsection{2.2 {}{} Validating User Hypotheses}

To evaluate each slice discovery algorithm, our study focuses on the human-centered task of \emph{hypothesis formation}, where a human describes in words where they believe the model underperforms.
In this section, we formalize this task and discuss our approach for validating users' hypotheses.

We define a ``\emph{hypothesis}'' as a \emph{user-generated text description} of a group where they believe that the model underperforms.
We follow past work and give as input to each slice discovery algorithm sets of datapoints $D$ that all share the same true class label \citep{singla2021understanding, plumb2023rigorous}.
In this setting, ``errors'' correspond to images where the model failed to detect the object.
Intuitively, a ``correct hypothesis'' should describe a group where the model has significantly lower accuracy at detecting the object.
If $\Phi$ denotes the set of all images in $D$ that belong to the group, we define the \emph{performance gap} $Gap(f, \Phi)$ of $f$ on $\Phi$ as
\begin{equation}
    \underbrace{\frac{1}{| D | } \sum_{(x, y) \in D } 1(f(x) = y)}_{\text{average accuracy on $D $}} - \underbrace{\frac{1}{| \Phi| } \sum_{(x, y) \in \Phi} 1(f(x) = y)}_{\text{average accuracy on $\Phi \subset D$}}
\label{eqn:gap}
\end{equation}

We define the correctness of the hypothesis ``$f$ underperforms on $\Phi$'' by thresholding the gap:

\begin{equation}
    \text{``$f$ underperforms on $\Phi$''} \Leftrightarrow Gap(f, \Phi) > \tau
\label{eq:correct-threshold}
\end{equation}

where threshold $\tau \geq 0$ is a hyperparameter that defines the minimum performance gap necessary for a hypothesis to be ``correct''.
For example, if we set $\tau = 0.2$, then a ``correct'' hypothesis describes a group where the model has 20\% worse accuracy on images that belong to the group, compared to all images that belong to its class.

One major challenge is that in many settings, we do not have access to $\Phi$, the complete set of images in $D$ that match each hypothesis.
Unfortunately, obtaining $\Phi$ is often difficult and expensive.
For example, recall the scenario where a practitioner hypothesizes that her model underperforms at detecting stop signs for ``photos taken in snowy environments''.
In this setting, the practitioner a-priori does not know which images in the class $D$ match her hypothesis.
She could manually review all photos of stop signs and annotate whether they were taken in snowy environments, but this strategy is slow and inefficient.

In our study, we aim to validate the correctness of $180$ user-generated hypotheses.
Unfortunately, manually reviewing each image within each class $D$ to obtain the full set of images that match each hypothesis is difficult at scale.
Thus, we decided to approximate the model's performance gap for each hypothesis using a sample $\hat{\Phi}$ of images that match each hypothesis, where $\hat{\Phi} \subset \Phi$. 
We provide an extended description of our approximation strategy in Appendix B, and summarize key steps below.

\paragraph{Approximating the performance gap}
Our goal is to find a sample of images $\hat{\Phi}$ from the class $D$ that match each hypothesis.
To do this, we follow past work \citep{gao2022adaptive} and rank candidate images using their CLIP similarity score \citep{radford2021learning} to the hypothesis text description.
When relevant, we performed minimal prompt engineering to users' hypotheses to increase the quality of the similarity scores.
Following \citet{vendrow2023dataset} and \citet{gao2022adaptive}, the final edited prompts began with the phrase, \emph{``a photo of a [class] and [...]''}.\footnote{Like past work, we use different prompt templates for hypotheses that do not fit the general template.  For example, we modify the template for artistic styles (\ie{} ``a \emph{greyscale} photo of an airplane'').  }

Next, we inspected a sample of the images most similar to each hypothesis, and manually labeled up to the first $40$ images that matched the  hypothesis.
In an effort to make our labeling process consistent and reproducible, we created a labeling guide inspired by \citet{shankar2020evaluating} describing the criteria we used to determine whether an image matched each hypothesis.
We provide an extended description of this labeling process in Appendix C.

For each class, we used the same CLIP retrieval strategy to obtain a sample $\hat{D}$ of images belonging to the class.  
We define each $\hat{D}$ as the $100$ images with the highest similarity score to prompt \emph{``a photo of [class]''}.  
Our final approximation of the performance gap calculates the difference in the model's accuracy for the group's sample $\hat{\Phi}$ vs. the class's sample $\hat{D}$.
We publicly release our labeling guide and the files we selected as matching each group description (\ie{} the $\hat{\Phi}$) on GitHub.\footnote{https://github.com/njohnson99/slice-discovery-human-eval}
We discuss several ablations to our approximation strategy in Appendix B.

%% file: text/hcomp-sections/related-work.tex
\section{Related Work}\label{sec:rw}

\paragraph{Automated Evaluations of Slice Discovery Tools}  
Automated evaluations of slice discovery tools fall into three categories.  
The first category of evaluations measure the quality of each slice by calculating its size or error rate \citep{d2021spotlight, singla2021understanding}, but fail to capture any notion of the slices' coherence.
The second category of evaluations compare the output slices to known groups where the model underperforms \citep{sohoni2020no, eyuboglu2022domino, plumb2023rigorous, wang2023error}.
One major limitation of this approach is that in practice, we often lack access to the full set of groups where a model underperforms.
As such, these works evaluate methods' ability to discover only the same subset of known errors, \ie{} groups that have already been annotated in well-studied benchmarks \citep{sohoni2020no}.
The final category of evaluations uses another ML model (in place of a human) to generate a natural language description of each slice \citep{eyuboglu2022domino, gao2022adaptive}.
However, recent work has shown that these ML models often generate nonsensical descriptions \citep{gao2022adaptive}.
Our study avoids limitations of past automated evaluations by instead running a human evaluation that asks people (rather than a separate ML model) to describe each slice.
Our hypothesis validation approach also does \emph{not} require access to the full set of true model errors.

\paragraph{Human Evaluations of Slice Discovery Tools}
Several works that introduce new slice discovery tools include qualitative evaluations performed by humans.
Many such works do not conduct a controlled user study.  
Instead, the authors themselves describe the semantic features shared by the top $3$, $5$, or $10$ images in each slice \citep{d2021spotlight, singla2021understanding, eyuboglu2022domino, wang2023error}.
To our knowledge, the majority of these evaluations do not validate that their descriptions accurately depict groups where the model does underperform.
Some works use the authors' ability to find a description that matches the majority of images in the slice as a proxy of the algorithms' utility.
For example, \citet{d2021spotlight} argue that their method is superior to a competitor because the latter's output slices appear to ``\emph{share little in common}''.

One of the most similar studies to ours is \citet{singla2021understanding}, where industry data scientists are asked to use slices output by an algorithm to develop hypotheses about where an image classifier fails.
However, their study does not validate the correctness of participants' hypotheses.
Further, they only evaluate a single slice discovery algorithm.

In contrast to past work, our work does \emph{not} assume that users' hypotheses are de facto correct depictions of true errors.
Instead, we validate each user-generated hypothesis by calculating the model's accuracy on a new sample of data.
Furthermore, we explicitly study whether output slices' coherence is a valid proxy for tools' utility.

\paragraph{Behavioral Understanding of ML}
A growing number of works propose tools to facilitate stakeholders' understanding of model behavior.
Specifically, our work contributes to the growing body of empirical studies that ask users to form hypotheses about where a model underperforms 
\citep{singla2021understanding, cabrera2022what, suresh2023kaleidoscope, moore2023failure}.
However, relatively few existing studies validate the correctness of users' hypotheses \citep{wu-etal-2019-errudite, cabrera2021discovering, gao2022adaptive}.
To our knowledge, ours is the first study to evaluate whether slice discovery tools help users form \emph{correct} hypotheses of model behavior.

%% file: text/hcomp-sections/methodology.tex
\section{Experimental Design}

We conduct a controlled user study motivated by three high-level goals, under which we organize our hypotheses:
\begin{enumerate}
    \item Our first goal is to \emph{benchmark} two state-of-the-art slice discovery tools (``conditions''), \textsc{Domino} and \textsc{PlaneSpot}, against a naive baseline.
    This comparison provides an important sanity check that these tools may offer users some benefit over a naive workflow of inspecting a random sample of errors.

    We were specifically interested in comparing three \emph{measures} associated with each hypothesis, across conditions:  how well the user's description of the slice (hypothesis) captures a group where the model actually does underperform (``correctness''), how difficult it was for the user to describe the slice, and how many images in the slice match the user's hypothesis:
    
    \textbf{H1}.  A greater proportion of users' hypotheses corresponding to slices output by slice discovery algorithms will be correct, when compared to hypotheses corresponding to slices output by the naive baseline.

    \textbf{H2}.  Users will rate slices output by slice discovery algorithms as easier to describe, when compared to slices output by the naive baseline.

    \textbf{H3}.  Users will select more images as ``matching'' their hypothesis for slices output by slice discovery algorithms, when compared to slices output by the naive baseline.
    
    \item Our second goal is to characterize the relationship \emph{between} these measures for the slices output by state-of-the-art algorithms (\textsc{Domino} and \textsc{PlaneSpot}).

    As discussed in Section \ref{sec:rw}, past evaluations use measures of slice \emph{coherence}, such as whether the majority of images share a common description, as a proxy of algorithms' utility.
    However, this assumption overlooks that just because many images in the slice match some description, does not mean that the model underperforms on \emph{all} such images.
    To examine this assumption, we explicitly study the relationship between the number of images in the slice that match the users' hypothesis (out of $20$ shown), and its correctness. We note that the number of matching images is just one possible heuristic to capture the coherence of the slice, and that there may be other valid ways to measure slice coherence.

    \textbf{H4}.  The number of images in the slice that match a user's hypothesis (a measure of slice coherence) does not predict whether their hypothesis is correct.
    
    \item Our final goal is to explore how humans make sense of the slices output by state-of-the-art algorithms.  
    
    One assumption shared by past work is that each slice corresponds to a \emph{single unique group} where the model underperforms.
    However, in pilot studies we observed that different users arrived at different conclusions about model behavior, even when shown the exact same information (details in Appendix D).
    
    Our final hypothesis studies different users' consistency with each other when shown the same slice.
    We ask annotators to label whether users' hypotheses are synonymous (\ie{} may differ syntactically, but describe the same group of images), or distinct.

    \textbf{H5}.  Different users will write down distinct hypotheses when presented with the same slice.
    
\end{enumerate}

\begin{figure*}[h]
\centering
\includegraphics[width=\linewidth]{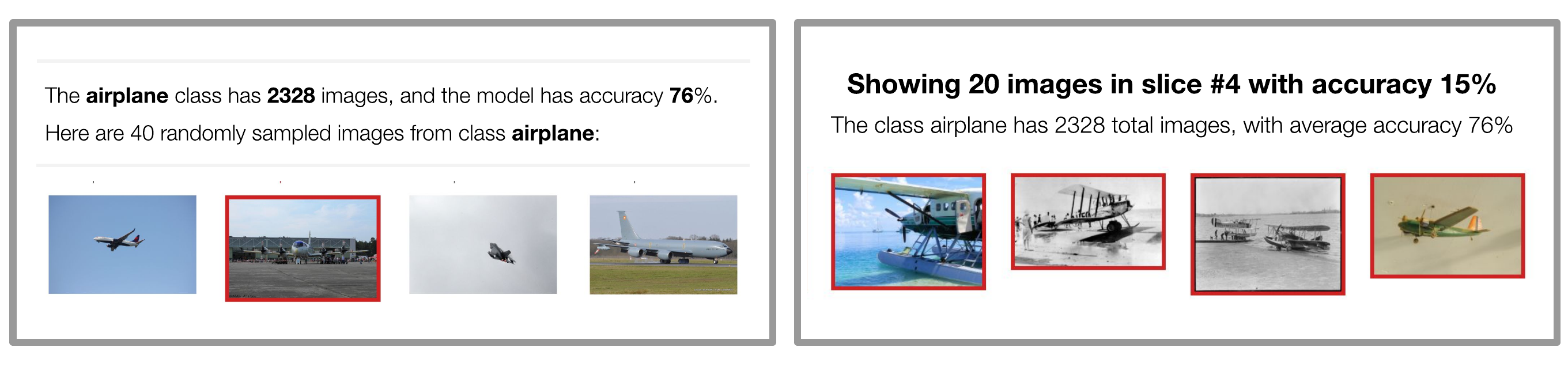}
\caption{UI screenshots of the class overview (Left) and slice overview (Right). Errors (where the model failed to detect the object) have red borders. (Left, Top) The class overview shows the total number of images in the test set that belong to the class and the model's average accuracy on these images.  (Left, Bottom) A random sample of $40$ images from the test set. (Right, Top) The slice overview shows model's average accuracy on the top-$20$ images belonging to the slice and the entire test set.  (Right, Bottom) The top-$20$ ordered images that belong to the slice.}
\label{fig:ux-main}
\end{figure*}

\subsection{4.1 {}{} Study Design}\label{sec:design}

\subsubsection{Domain \& Model}
To study whether slices can help humans understand model behavior, we must first select a domain, prediction task, and model.
We used data from MS-COCO (``COCO'') \citep{lin2014microsoft}, a large object detection dataset,
for its accessibility to a wide audience.  
COCO contains photos with $91$ different object types ``\emph{that would be easily recognizable by a four-year-old}'' \citep{lin2014microsoft} in everyday natural scenes.
COCO is a multi-label classification task where a single image can have several objects present.
We defined a custom 15\%-10\%-75\% train-validation-test split so that we could use a larger held-out test set as input to the slice discovery algorithms.
We fine-tuned a pretrained ResNet-18 model using the training set (details in Appendix E), performed model selection using the validation set, and ran the slice discovery algorithms on the held-out test set.
Because we aim to evaluate how well slice discovery tools can detect naturally occurring errors, we did not modify the dataset or model training process to synthetically induce specific errors.

\begin{figure}[H]
    \centering
    \includegraphics[width=0.95\linewidth]{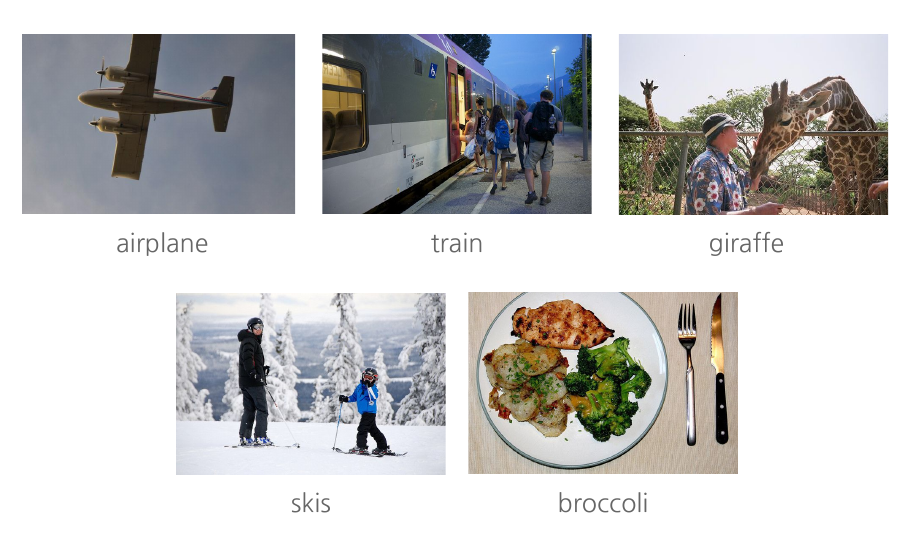}
\caption{Example photos from the $5$ selected COCO object classes \citep{lin2014microsoft}.}
\label{fig:coco-examples}
\end{figure}

\subsubsection{Computing Slices}\label{sec:experiments-sdms}
We generated $60$ total slices ($20$ slices per each of the $3$ algorithms) to show users. 
Rather than run each slice discovery algorithm on the entire COCO test set (which contains $91$ object types), we followed past evaluations \citep{gao2022adaptive, plumb2023rigorous} and ran each algorithm on subsets of the data that all have the same object.
Thus, each hypothesis describes a group where the model has low \emph{recall} (\ie{} failed to detect the object).
We used each slice discovery algorithm to return the top $k = 4$ slices for $5$ different objects (Figure \ref{fig:experimental-design}, Top), chosen randomly from a list of candidate objects where the model had at least 50\% recall: \texttt{airplane}, \texttt{train}, \texttt{giraffe}, \texttt{skis}, and \texttt{broccoli}.

\subsubsection{Participants \& Recruitment}  
We elicited $12$ hypotheses each from $15$ total subjects, for a total of $180$ user-generated hypotheses.
We recruited subjects that had self-reported ``intermediate knowledge in machine learning (ML) or computer vision (CV)'' \ie{} had taken a graduate-level course or had practical work experience in ML, AI, or CV, using university mailing lists.
We recruited participants with ML expertise (rather than task-specific expertise) because most existing slice discovery tools were created to be used by model developers.
All participants were students enrolled in a full-time degree program in computer science.
Participants were compensated with a \$20 gift card, and reported spending $30$ (min) to $55$ (max) minutes participating.

\subsection{4.2 {}{}  Study Procedure}
The study was approved by an Institutional Review Board (IRB) process and was conducted asynchronously online.
Participation was voluntary and users were shown a consent form before participating.  
We began each study by presenting the user with a description of the study task and walk-through of the study interface.
After completing the walk-through, the user was shown information about $12$ different slices belonging to $3$ different object classes.
The user completed a questionnaire that asked them to formulate a behavioral hypothesis for each slice.
We detail each phase of the study procedure below.

\paragraph{Instructions}  
In the study walk-through, we introduced the user to the object detection task by showing them example images and model predictions from a randomly selected class (``tennis racket'').
We defined and motivated the slice discovery task by listing several reasons why one may wish to discover groups where a model underperforms.
We provide screenshots of the task instructions in Appendix F.

\paragraph{Class Overview}  Each user was asked to form hypotheses for slices corresponding to $3$ different object classes.
For each class, the user was first shown a \emph{class overview} (Figure \ref{fig:ux-main}, Left) with information about and example images belonging to the class.
We presented the class overview before presenting the slices to give the user basic context about the COCO dataset (\ie{} examples to illustrate the variety of images that belong to each class) that stakeholders in practice would already have for their domain.

\paragraph{Slice Overview \& Questionnaire}
For each slice, the user was shown a \emph{slice overview} (Figure \ref{fig:ux-main}, Right) that displays the model's performance on the top $20$ images that belong to the slice.
The user was asked to use this information to complete the \emph{slice questionnaire}, which asked them to (1) write down a behavioral hypothesis of the underperforming group corresponding to the slice, (2) select all images in the slice that belong to this group, and (3) rate how difficult it was to describe the slice. We provide the complete questionnaire in Appendix G.

\emph{Visualizing each slice.} We controlled for how we visualized the model's performance on each slice by presenting the top-$20$ images only for all conditions.
We designed our slice overview to be as similar as possible to how past work presents individual slices \citep{d2021spotlight, eyuboglu2022domino}.
We limit the number of images shown to $20$ because showing more images may cause information overload and increase the time required to complete each questionnaire.

\emph{Eliciting a hypothesis for each slice.} 
Past works assume that \emph{all images} in each computed slice correspond directly to a \emph{single} group where the model underperforms \citep{d2021spotlight, eyuboglu2022domino, plumb2023rigorous}.
We explicitly evaluate this assumption and ask users to form a single behavioral hypothesis for each slice.
Specifically, we instruct users to \emph{describe} the slice to the best of their ability by writing a group description that matches as many images in the slice as possible.
Users are told that some slices may be noisy or incoherent, and that in some cases it may be difficult to find a single description that matches all of the images in the slice.
We provide users with further guidance and example hypotheses detailed in Appendix F.

\vspace{-3mm}
\subsection{4.3 {}{} Experimental Design}\label{sec:exp-design}
\vspace{-1mm}

\begin{figure}[h]
    \centering
    \includegraphics[width=\linewidth]{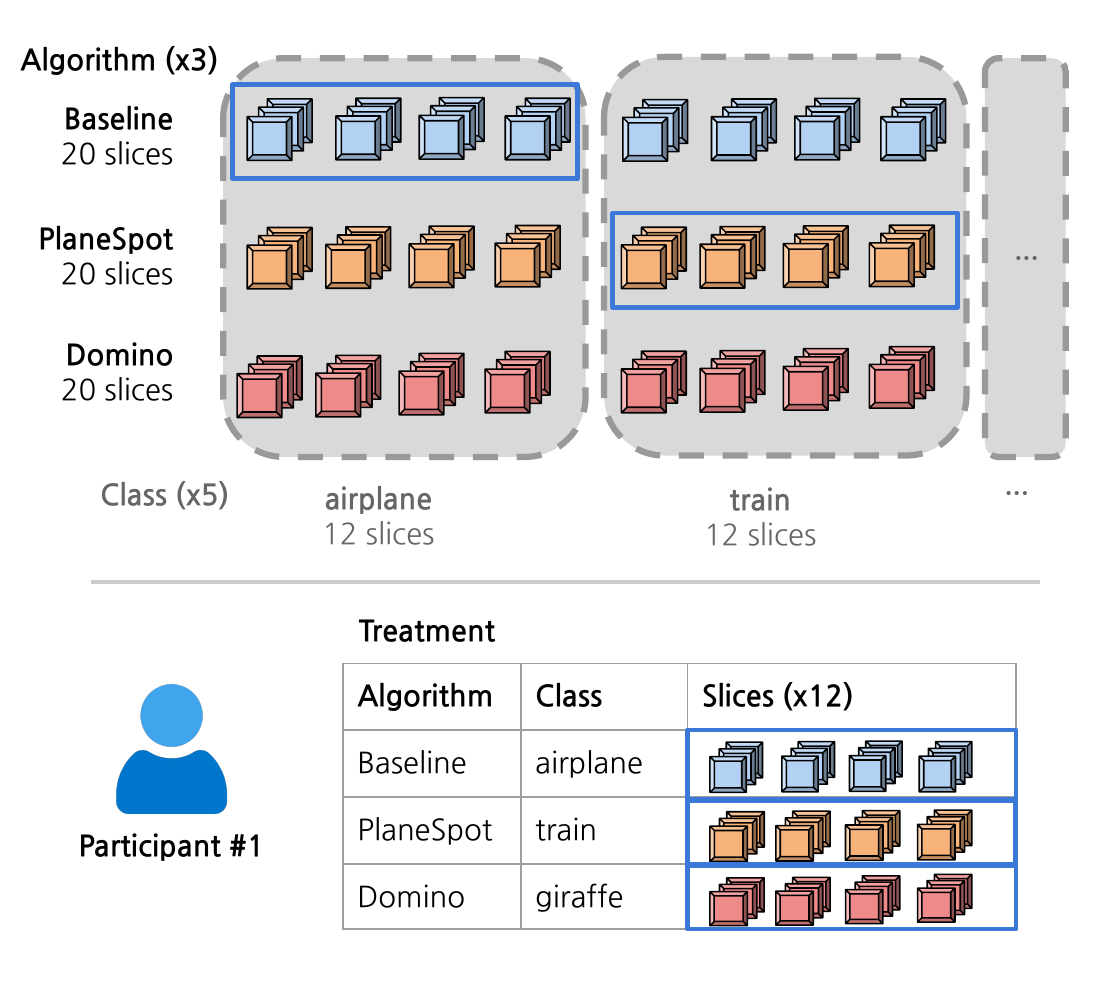}
\caption{Experimental Setup. (Top) We collect users' hypotheses for $60$ total slices.  For each of the $3$ algorithms (rows), and for each of the $5$ classes (columns), we compute the top-$4$ slices. 
We show $2$ out of $5$ classes (and $24$ out of $60$ total slices) in the figure due to space constraints.
(Bottom) Each study participant was shown $12$ total slices output by $3$ different algorithms, and saw slices corresponding to a different class for each algorithm.  For example, Participant \#1 was asked to develop hypotheses for the top-$4$ slices output by \textsc{PlaneSpot} for the \texttt{train} class.  The blue boxes on the top panel highlight the slices shown to Participant \#1.}
\label{fig:experimental-design}
\end{figure}

We used a within-subjects design shown in Figure \ref{fig:experimental-design}, where each participant was randomly assigned to $3$ (out of $5$ candidate) object classes.
All participants were presented with $12$ total slices, and saw slices output by all $3$ algorithms.
The $3$ algorithm conditions were presented to participants in a random order to control for learning effects.
We showed each of the $60$ total slices to $3$ different participants, collecting $3$ hypotheses per each slice.
Users were blinded to the study condition when completing each questionnaire: while they were informed that each slice was computed by an algorithm, they were not given any information about \emph{which algorithm} was used to compute each slice.

\vspace{-1mm}
\subsection{4.4 {}{} Metrics \& Analyses}
To evaluate \textbf{H1} (whether the average number of correct hypotheses varies across conditions), we exclude hypotheses where we failed to find a sufficiently large sample of matching images to approximate the performance gap.
Applying this criteria, we retain $136$ out of the original $180$ hypotheses (76\%) where we found a sample of at least $15$ matching images.
To evaluate \textbf{H2} and \textbf{H3}, we calculate each measure using all $180$ of the original hypotheses.

To test for statistically significant differences between the three algorithm conditions, we ran ANOVA tests with Tukey post-hoc tests for multiple comparisons to compare the proportion of correct hypotheses (\textbf{H1}) and average number of matching images (\textbf{H3}).  We ran Mann-Whitey tests with Bonferroni corrections to compare the Likert-scale self-reported difficulty of describing each slice (\textbf{H2}).

To evaluate \textbf{H4} and \textbf{H5}, we retained only the hypotheses that correspond to slices output by slice discovery algorithms, and excluded slices output by the baseline condition.

To determine if coherence (\ie{} the number of images in the slice that match a hypothesis) is an appropriate proxy of the hypothesis's  correctness (\textbf{H4}), we ran two Spearman's rank correlation tests.
For both tests, the independent variable is the number of matching images.
We tried two dependent variables: the value of the approximate performance gap for the hypothesis (defined in Equation \ref{eqn:gap}), and an indicator for hypothesis correctness using performance gap threshold $\tau = 0.2$.

To determine whether two different users' hypotheses are equivalent or distinct (\textbf{H5}), we asked two annotators to label groups of hypotheses that are synonymous (\ie{} describe the same group of images).
For each slice, we then use these groups to count the number of distinct user hypotheses (where each group of synonymous hypotheses only counts as a single ``distinct hypothesis'').
We discuss this process of determining whether two hypotheses are distinct or synonymous in detail in Appendix H.

%% file: text/hcomp-sections/results.tex
\begin{figure*}
\centering
\begin{minipage}{0.45\textwidth}
  \centering
  \includegraphics[width=\linewidth]{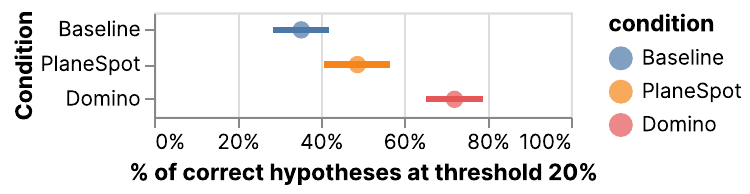}
  \captionof{figure}{Hypothesis Correctness. The percentage of hypotheses per condition that are ``correct'' using a performance gap threshold $\tau = 20\%$ with standard error bars.  Percentages are calculated for the subset of hypotheses that we have a sufficiently large number of examples (\ie{} at least $15$ matching images) to approximate the performance gap.  We find that a greater proportion of users' hypotheses from the \textsc{PlaneSpot} and \textsc{Domino} conditions are correct relative to the \textsc{Baseline} condition.}
   \label{fig:correctness}
\end{minipage}%
\qquad \qquad
\begin{minipage}{0.45\textwidth}
  \centering
  \includegraphics[width=1\linewidth]{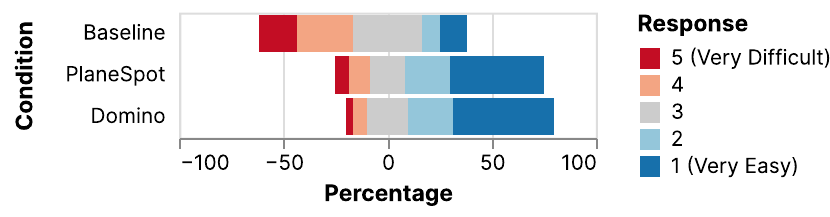}
  \captionof{figure}{Self-Reported Difficulty.  A diverging stacked bar chart centered around the netural response of users' self-reported difficulty of describing each slice (on a five-point Likert Scale), stratified by each condition (slicing algorithm).  We find that users are significantly more likely to rate slices output by the \textsc{PlaneSpot} and \textsc{Domino} conditions as being easy to describe.}
  \label{fig:reported-difficulty}
\end{minipage}
\end{figure*}

\begin{figure*}
\centering
\begin{minipage}{0.45\textwidth}
  \centering
   \includegraphics[width=0.7\linewidth]{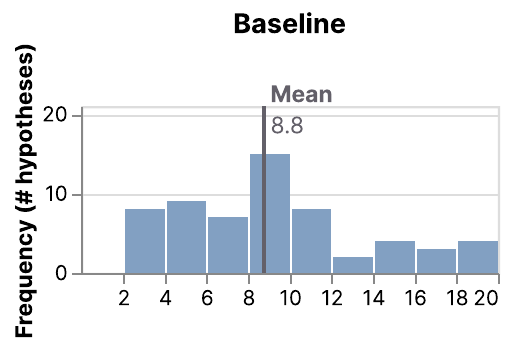}
  \includegraphics[width=0.7\linewidth]{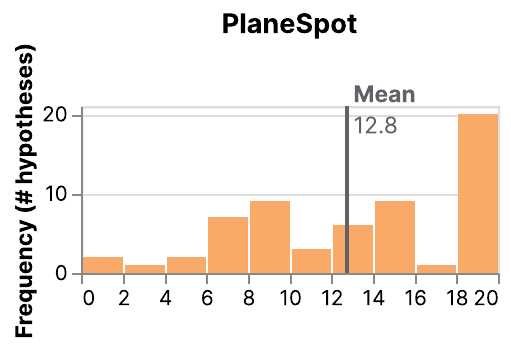}
  \includegraphics[width=0.7\linewidth]{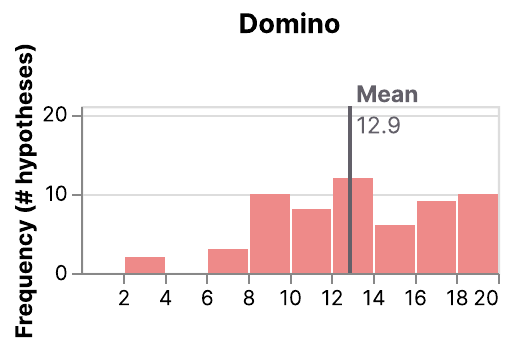}
  \captionof{figure}{Number of Matching Images. Histograms of the number of images in each slice that match the user's hypothesis (out of the top $20$), stratified by each condition (slicing algorithm).  The mean number of matching images for each condition is denoted by a vertical line.  We find that on average, users selected more images as ``matching'' their hypothesis for slices output by \textsc{PlaneSpot} and \textsc{Domino}.}
   \label{fig:matching-images}
\end{minipage}%
\qquad \qquad
\begin{minipage}{0.45\textwidth}
  \centering
  \includegraphics[width=\linewidth]{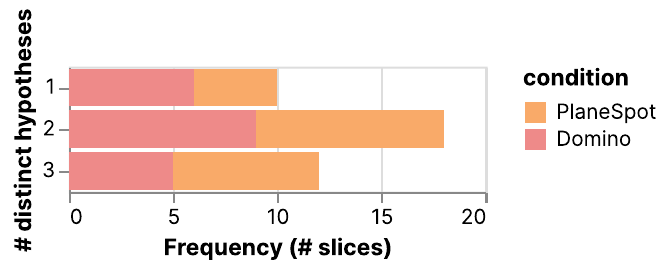}
  \captionof{figure}{User Consistency Barplot. A stacked bar plot visualizing the number of slices from the \textsc{Domino} and \textsc{PlaneSpot} conditions ($x$-axis) that have $1$, $2$, and $3$ distinct hypotheses (of the three different users' hypotheses) ($y$-axis).  We find that at least two out of three users write down distinct hypotheses for the majority (75\%) of \textsc{Domino} and \textsc{PlaneSpot} slices.}
  \label{fig:user-consistency}

  \vspace{25pt}
    \includegraphics[width=\linewidth]{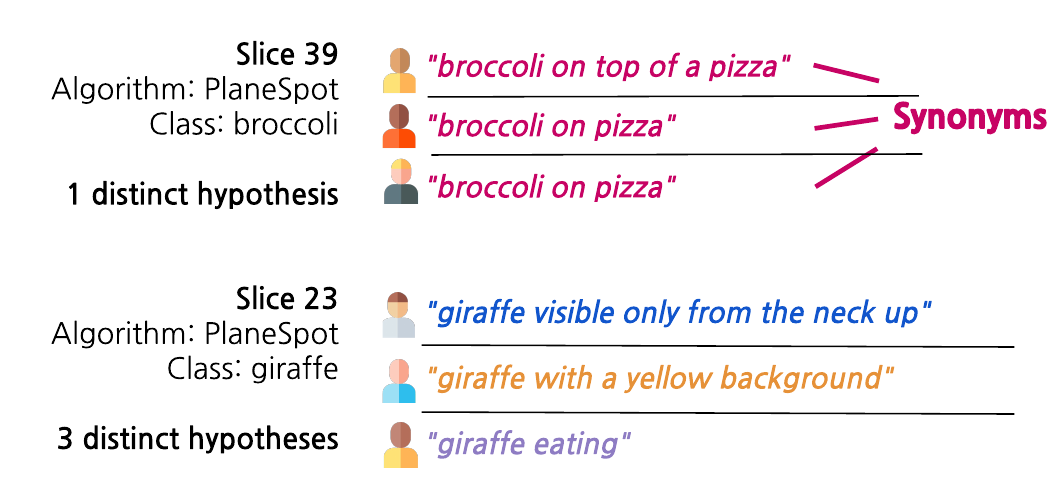}
  \captionof{figure}{User Consistency Examples. Users' hypotheses for two example slices output by \textsc{PlaneSpot}.  For the top slice (\#39), all three users wrote down synonymous hypotheses that the model underperforms on images of ``\emph{broccoli on pizza}''.  In contrast, for the bottom slice (\#23), the three users wrote down distinct hypotheses of where the model underperforms.  }
  \label{fig:consistency-examples}

\end{minipage}

\end{figure*}

\section{Results}

We used a $p$-value of $0.05$ as our cutoff for significance.
We present additional details for each statistical test (\eg{} test statistics) and additional analyses in Appendix I.

\paragraph{Correctness} For each condition, we calculated the proportion of hypotheses that are ``correct'' using a performance gap threshold of at least $20\%$.  
We found that $70$ out of $136$ total hypotheses (51\%) are correct (standard error 4.3\%).  When we stratify by condition, 35\% of \textsc{Baseline}, 49\% of \textsc{PlaneSpot}, and 73\% of \textsc{Domino} hypotheses are correct, with standard errors 6.7\%, 7.9\%, and 6.8\% respectively  (Figure \ref{fig:correctness}).
We found a statistically significant difference between the \textsc{Baseline} and \textsc{Domino} conditions only $(p < 0.001)$, partially supporting \textbf{H1}.

We present additional results where we ablate the performance gap threshold $\tau$ in Appendix I.1.
In summary, we found that a consistent trend (that \textsc{Domino} outperforms \textsc{PlaneSpot}, which outperforms \textsc{Baseline}) holds for all variations tried.
\textsc{Domino} consistently significantly outperformed the \textsc{Baseline} condition; however, there was not a statistically significant difference between the other conditions for the majority of thresholds.

\paragraph{Number of matching images}
Overall, we observed that users selected more images as ``matching'' their hypothesis for slices output by slice discovery algorithms relative to the naive baseline, supporting \textbf{H2}.
Figure \ref{fig:matching-images} shows a histogram of the empirical distribution of matching images for each condition.
On average, users selected $12.9$ and $12.8$ out of the $20$ displayed images in the slice as ``matching'' their hypotheses for the \textsc{Domino}
and \textsc{PlaneSpot} conditions respectively (standard errors $0.57$ and $0.71$), vs. $8.8$ out of the $20$ images (standard error $0.60$) for the \textsc{Baseline} hypotheses.
We observed significant pairwise differences between the two slice discovery conditions vs. the baseline condition (with $p < 0.0001$ for both \textsc{Domino} and \textsc{PlaneSpot}), and no significant difference between the two slice discovery conditions.
However, we note that while the \emph{average} number of matching images is higher for both slice discovery conditions, there is still high \emph{variance} across hypotheses.
For example, 30\% of hypotheses from the \textsc{Domino} or \textsc{PlaneSpot} conditions had $< 10$ matching images.

\paragraph{Self-reported difficulty}
We observed that users rated slices output by the slice discovery algorithms as \emph{easier to describe} compared to slices output by the naive baseline condition, supporting \textbf{H3}. 
Figure \ref{fig:reported-difficulty} shows the distribution of Likert-scale ratings for each condition. 
We observed significant pairwise differences between the two slice discovery conditions vs. the baseline ($p < 0.0001$), and no significant difference between the two slice discovery conditions.  

\paragraph{Does coherence imply correctness?}
We found \emph{no significant association} between the number of images that match each hypothesis and its correctness, supporting \textbf{H4}. 
For both dependent variables, the number of matching images is only weakly correlated with the hypothesis's correctness, with correlation coefficients $r_s = 0.08$ with $p = 0.4712$ for the value of the performance gap, and $r_s = 0.06$ with $p = 0.5996$ for the correctness indicator.
When we examined the hypotheses that had the highest number of matching images (at least 15 out of 20), only 60\% of these hypotheses were correct, a rate comparable to the overall base rate of 61\% accuracy for all hypotheses corresponding to slices output by either \textsc{Domino} or \textsc{PlaneSpot}.

\paragraph{(In)consistency across users}
Figure \ref{fig:user-consistency} shows the number of distinct hypotheses for each of the $40$ slices output by either \textsc{PlaneSpot} or \textsc{Domino}, and Figure \ref{fig:consistency-examples} shows users' hypotheses for two example slices output by \textsc{PlaneSpot}.
We found that all three users wrote down consistent (\ie{} synonymous) hypotheses for only a small minority (25\%) of slices, supporting \textbf{H5}.
In contrast, at least $1$ of the $3$ users wrote down a hypothesis that differed from other users for the majority (75\%) of slices, and all three users disagreed for 30\% of slices.

%% file: text/hcomp-sections/discussion.tex
\vspace{-3mm}
\section{Discussion}

We share several implications of our experimental results (Section 5.1) and present design opportunities for future tools to help users understand where their model underperforms (Section 5.2).

\subsection{6.1 {}{} Interpreting the Experimental Results}

We return to our three high-level study goals to place our results in conversation with past work on slice discovery.

\paragraph{Benchmarking existing tools} 
Overall, we found significant differences across measures between the \textsc{Domino} and \textsc{PlaneSpot} conditions, relative to the naive baseline algorithm.
Specifically, while \textsc{Domino} and \textsc{PlaneSpot} slices had a significantly higher average number of matching images and were significantly easier to describe relative to the baseline condition, supporting \textbf{H2} and \textbf{H3}, only \textsc{Domino} had a statistically significant difference in the proportion of correct hypotheses.
While \textsc{PlaneSpot} had a higher proportion of correct hypotheses than the \textsc{Baseline} condition, the difference was not statistically significant for the majority of performance gap thresholds $\tau$ (Appendix I.1).

In summary, our results provide preliminary evidence that existing tools may offer some benefit relative to a naive workflow of inspecting and trying to make sense of a random sample of errors.
However, the slices output by existing tools do not \emph{always} help users form correct hypotheses.
For example, only 49\% of users' hypotheses corresponding to slices output by \textsc{PlaneSpot} described groups where the model performed at least 20\% worse.
Thus, existing tools do have the potential to mislead users to develop false beliefs if used as proposed by past work.

\paragraph{Coherence does not imply correctness} 
Our study calls into question previous assumptions about slice \emph{coherence} and hypothesis \emph{correctness} (\textbf{H4}).
A number of works have implied that if the user can describe all the images in a slice, that the model underperforms on all images that match their description.
Our finding that there is no significant association between slice coherence and hypothesis correctness shows that this assumption is misleading.

While it may seem counter-intuitive, we are \emph{not} surprised that existing slice discovery tools tend to underrepresent the model's performance on coherent groups.
We hypothesize that this behavior is a feature (rather than a bug) resulting from the way that existing tools are designed to return high-error subsets.
Existing tools may be incentivised to output misleading slices that contain a misrepresentative sample of high-error images \emph{within} each group, even in cases where the model doesn't actually perform worse on the \emph{entire} group.
Evaluations that only consider the coherence of the output slices fail to account for whether the slice is a representative sample of the model's performance on the entire group.

This finding has significant implications for researchers, who should center hypothesis correctness (rather than slice coherence) when evaluating their tools.
But perhaps more importantly, this finding is significant for \emph{users} of these tools.
While present-day tools can help users form hypotheses, users should take caution to know that their hypotheses, no matter how aligned with the slices they may be, may not necessarily be correct.
Thus, validating behavioral hypotheses is not only important for researchers, but also for users.

\paragraph{Variance across users}  
Our finding that different users form different hypotheses when down the same slice points to the points to the complexity of hypothesis formation as a \emph{human-centered} task.
This finding contradicts the dominant assumption from ML research that all users can easily identify the semantic features shared by a subset of data, or that users will simply ``know it when they see it''.
But, disagreement among users whom are asked to complete the same task has been long-demonstrated and well-studied in the field of \emph{crowdsourcing} \citep{callison-burch-2009-fast, chang2017revolt}.
In practice, different stakeholders may bring different prior knowledge that may cause them to develop different hypotheses when shown the same slice.
If handled thoughtfully, however, this variance across users' hypotheses may actually serve as a strength: for example, a team of stakeholders could discuss several candidate hypotheses for a given slice \citep{drapeau2015microtalk}.
More generally, we believe that the field of slice discovery has much to learn from related work on crowdsourcing and annotation.

\subsection{6.2 {}{} Design Opportunities}

Our findings point to several design opportunities for ML and HCI researchers.  
We highlight a few exciting directions for future work below.

\paragraph{Supporting hypothesis formation}  
In the hypothesis formation step, stakeholders make sense of a large amount of information (\eg{} the images belonging to each slice) to form hypotheses about model behavior.
However, the human factors that affect participants' hypothesis formation process have been understudied.

One dimension that has been neglected by past work is how the output slices should be \emph{visualized} and presented to users.
When asked if there was ``any other information that they were not shown that [they] believe would have helped [them] complete the questionnaire'', several users expressed a desire to see the model's performance on a larger set of examples beyond the top-$20$ images belonging to the slice.
One user wrote, ``\emph{it would have been helpful to retrieve other images from the dataset during slice labeling} (hypothesis formation)'' to ``\emph{see how good my label} (hypothesis) \emph{was}''.

We envision several alternative ways to present each slice beyond the static grid used in our slice overview.
For example, one could design novel interactive workflows that allow users to explore the model's performance on multiple slices simultaneously \citep{bertucci2022dendromap, cabrera2023zeno}, or compare the images within to images outside of each slice.
Designing an appropriate visualization that accounts for the cognitive load and biases of the user \citep{gajos2017influence, cabrera2022what} is an important direction for future work.
Further, the most helpful information to show the user may vary depending on their expertise, \eg{} whether the user is an ML developer or a domain expert.

\paragraph{Prioritizing hypothesis correctness}
Our study shows that slices that appear to be coherent can mislead users to develop incorrect beliefs about where their model underperforms.
Thus, we urge researchers to prioritize developing tools that help users form \emph{correct} hypotheses.
One problem we identified is that existing slice discovery tools are incentivised to output high-error samples that may not capture the full diversity of images that belong to each group.
Future work could study how to output slices that contain a more representative sample of each group.

\paragraph{Towards real-time hypothesis validation} Another promising direction is to develop tools that allow users to validate their own hypotheses in real time.
Such tools would enable users to iteratively refine their hypotheses (\ie{} explore multiple possible errors that could correspond to a slice), and serve as a sanity check before investing in expensive downstream actions based on false beliefs.
Most existing workflows to help users collect evidence to validate their hypotheses often prioritize retrieving examples that are most \emph{similar} to those that support the users' hypothesis \citep{cabrera2022what, gao2022adaptive, suresh2023kaleidoscope}.
In contrast, one potentially promising direction is to prioritize retrieving ``counter-evidence'': examples that contradict the users' hypothesis.
We hypothesize that counter-evidence may help combat stakeholders' potential confirmation bias and help them more quickly iterate on their hypotheses.

\paragraph{Interactive workflows for slice discovery}  
We encourage researchers to re-imagine where and how automation can help users discover underperforming groups.
Our finding that different users construct inconsistent descriptions of the same group of datapoints calls into question the workflow put forward in past work.
Existing slice discovery tools are meant to be run once, and then a stakeholder must make sense of the groups of datapoints they output.
One could imagine more bespoke and interactive workflows that better utilize stakeholder's domain knowledge to define coherent subsets of data, or leverage automation to help stakeholders refine their hypotheses.
As one example, a stakeholder could guide a clustering algorithm using their contextual understanding of semantic similarity \citep{rajani2022seal, cabrera2023zeno} rather than simply looking at the output clusters.

\subsection{6.3 {}{} Limitations}

The participants in our study were students with intermediate knowledge of machine learning, and we studied a model trained using data from a simple object detection task.
While much past work has developed tools intended to be used by model developers with ML expertise (such as the participants in our study), an emerging line of work has studied how to support users who do not have technical expertise, but \emph{do} have situated domain knowledge, in evaluating ML models \citep{suresh2023kaleidoscope}.
For example, clinicians \citep{gaube2023nontask} or content moderators \citep{suresh2023kaleidoscope} who have a deeper understanding of their data may interpret the output slices differently.
They may be able to use their domain knowledge to better characterize the semantic features shared by the examples in each slice.
Furthermore, in some domains, different stakeholders may disagree about desired model behavior, such as whether a comment should be moderated \citep{sap2022annotators}.
We encourage future work to further identify and examine these user-centered and context-specific dimensions of slice discovery.

While our work is an important step forward towards prioritizing hypothesis validation, we note that validating natural language hypotheses is difficult, and several open challenges remain.
Because retrieving all of the images that match each text description hypothesis is difficult and expensive, we follow \citet{gao2022adaptive} and approximate the model's performance on the group by retrieving a sample of matching images.
Unfortunately, the retrieved sample may not be representative of the model's performance on all images in the group; thus, we compare the model's performance to another retrieved sample to account for the implicit bias of our retrieval process (details in Appendix B).
For these reasons, our calculated performance gaps are only an \emph{approximation} of the model's performance on each group.
Despite these limitations, we found that all of our experimental findings were consistent across a range of ablations to our approximation strategy (Appendix I.4).
We believe that identifying inexpensive ways to retrieve a sample that matches a natural language hypothesis is an important direction for future work. 

Finally, our study focuses on the task of describing where the model underperforms \emph{in natural language}.
We asked users to describe the underperforming groups in words because doing so is a prerequisite for several downstream actions one might take to address the behavior, and for communicating about the behavior with a wider set of stakeholders.
We acknowledge that natural language hypotheses are imperfect for many use cases due to the implicit subjectivity or under-specification of natural language in some contexts.
For example, if a practitioner hypothesizes that her object detection model underperforms at detecting stop signs in photos where they are ``\emph{far away from the camera}'', determining which photos qualify as ``far away'' is under-specified from her description alone.
Developing a more formal, yet simultaneously accessible ``domain-specific language'' \citep{desai2015program, wu-etal-2019-errudite} for users' hypotheses is an open direction for future work.

%% file: text/hcomp-sections/conclusion.tex
\section{Conclusion}

While a growing number of works develop new slice discovery tools to help people discover where their model underperforms, there has been little evaluation of if, and how, humans can make sense of their output.
In our controlled user study with $15$ participants, we found preliminary evidence that existing tools may offer some benefit relative to a naive baseline of examining a random sample of errors.
Our results also challenge several dominant assumptions shared by past work on slice discovery.

First, coherence of the output slices does \emph{not} imply that users can form correct behavioral hypotheses.
Our results indicate that being able to identify a description that matches the majority of images in the slice does \emph{not} mean that the model underperforms on all such images.
This finding has important consequences for evaluation and everyday use of existing slice discovery tools.
We caution researchers away from using the number of images that match a common description as a measure of their algorithms' utility.

Second, we found that different users form different hypotheses when shown the same slice, which highlights the under-explored complexity of the hypothesis formation step.
Future work can consider alternative visualizations to help users make sense of the semantic features shared by the images in each slice, or make productive use of user disagreement to consider a wider range of possible model errors.

Our findings point to user needs and design opportunities to better support stakeholders as they form and validate hypotheses of model behavior.
More broadly, we hope that our work is a first step towards centering users when designing and evaluating new tools for slice discovery.

%% file: text/hcomp-sections/acknowledgements.tex
\newpage
\section*{Acknowledgments}

We thank our study participants and annotators who made our research possible.
We thank Sherry Tongshuang Wu, Donald Bertucci, Valerie Chen, Vijay Viswanathan, Jennifer Hsia, Katelyn Morrison, and Nupoor Gandhi for helpful feedback and discussions.
We also thank the reviewers at HCOMP 2023 and the ICML Second Workshop on Spurious Correlations, Invariance, and Stability for their suggestions that improved our paper.
This work was supported in part by the National Science Foundation grants IIS1705121, IIS1838017, IIS2046613, IIS2112471, and funding from Meta, Morgan Stanley, Amazon, and Google. 
Any opinions, findings and conclusions or recommendations expressed in this material are those of the author(s) and do not necessarily reflect the views of any of these funding agencies.

%% file: text/hcomp-sections/appendix.tex
\onecolumn
\newpage
\section{Slice Discovery Algorithms: Extended Descriptions}\label{apdx:sdms-extended}

\textsc{Domino} \citep{eyuboglu2022domino} works by first embedding each image using a multi-modal embedding.
We use CLIP \citep{radford2021learning} for its demonstrated performance on natural images. 
Next, \textsc{Domino} fits an error-aware Gaussian Mixture Model to model the input embeddings, model predictions, and true labels for each datapoint.  \textsc{Domino} outputs the top-$k$ mixture components with the largest discrepancy between the model's predictions and true labels.
We ran \textsc{Domino} using the default hyper-parameters used in their experiments ($\gamma = 10$).

\textsc{PlaneSpot} \citep{plumb2023rigorous} differs from \citet{eyuboglu2022domino} in two ways: 
First, \textsc{PlaneSpot} uses the final hidden layer activations of the neural network model that we aim to form hypotheses about (instead of a separate pretrained embedding).  
Second, while \textsc{PlaneSpot} also fits an error-aware Mixture Model, it does so by appending the model's predicted confidence to the model's embedding. 
\textsc{PlaneSpot} outputs the top-$k$ mixture components that have the largest product of their error rate and number of errors to prioritize large and high-error slices. 
We ran \textsc{PlaneSpot} using the default hyper-parameters used in their experiments (\ie{} running \texttt{scvis} using all default hyper-parameters from their python package, and $w = 0.025$).

Our \textsc{Baseline} algorithm randomly samples from the subset of misclassified images, without replacement.
To return $k$ slices that each have $m$ images, we randomly sample $m$ images without replacement from the set of misclassified images that have yet to be added to a slice.

Within each slice output by a slice discovery method, the datapoints are ordered by their component-conditional likelihood, where the ``most likely'' datapoints are thought to be the ``most representative'' of the semantic features that unify the slice \citep{eyuboglu2022domino}. 
The datapoints within each slice output by the \textsc{Baseline} algorithm are ordered arbitrarily.

\newpage
\section{Image Retrieval \& Approximation Strategy: Extended}\label{apdx:clip-retrieval}

Our goal is to retrieve a sample $\hat{\Phi}$ of images that match a text description hypothesis $t$, given a dataset of candidate images $D$.  At a high level, our approximation strategy has three steps:

\begin{enumerate}
    \item Use CLIP to compute a similarity score between the text description $t$, and every image in $D$.

    As proposed by the original paper \citep{radford2021learning}, we interpret the \emph{cosine similarity} between each (normalized) text description and image embedding as a measure of the semantic similarity between them.
    \item Manually inspect a sample of the top $80$ most similar images, and add up to the first $40$ images that match the description as belonging to $\hat{\Phi}$.  (This step is detailed in Appendix C.)
    \item Compare the model's accuracy on $\hat{\Phi}$, to the model's accuracy on another sample $\hat{D}$.
\end{enumerate}

\paragraph{Ablations}

We tried three versions of the above approximation strategy.  The first version (and its results) was presented in the main text.  We detail two alternative versions to highlight several subtleties that we discovered when considering how to best validate users' hypotheses.  

In summary, ideally we would retrieve a ``representative sample'' of images $\hat{\Phi}$ that is drawn randomly from $\Phi$, so that the model's performance on $\hat{\Phi}$ is an accurate representation of the model's performance on the entire group $\Phi$.  
However, we noticed that small changes to our CLIP retrieval strategy qualitatively appeared to influence the types of images that were retrieved.
Despite this variance, we found that all of our research hypotheses held across all retrieval strategies.
We highlight some of the subtleties we discovered with CLIP retrieval here to share with the research community, as CLIP image retrieval is becoming increasingly common in model debugging work.

\paragraph{Version \#1: Compare two samples retrieved by CLIP}

In Version \#1, we use the above strategy with no modifications.  For the final Step \#3, we compare the model's accuracy on $\hat{\Phi}$ to the model's accuracy on the top-100 images that are most similar to the prompt, ``\emph{a photo of [class]}'', \eg{} ``\emph{a photo of broccoli}''.

We chose to compare the model's performance to a sample $\hat{D}$ (rather than all of the images belonging to the class) to control for the implicit bias of CLIP's retrieval algorithm.
Table \ref{table:class-sample-accs} shows the model's average performance on the $100$ most similar images to the class prompt, vs. all of the images in each class.
With the exception of \texttt{skis}, for the majority of classes, the model performs much \emph{better} on the $100$ most similar images to the class prompt, especially for the \texttt{broccoli} class.
We hypothesized that this occurs because CLIP is more likely to rate images where the class is is an iconic view (\ie{} is not occluded) as being more similar to the class prompt.
Thus, because this implicit bias is likely reflected in the sample $\hat{\Phi}$ (\ie{} the true class may be more likely for an ML model to detect in the sample retrieved using CLIP), we decided to ``control'' for this implicit bias by comparing the model's performance to another sample retrieved using a similar prompt with the class name.

\begin{table}[h!]
\centering
\begin{tabular}{ c | c | c | c}
Class & \textbf{CLIP Sample} & Overall & $\Delta$\\ \hline
\texttt{airplane} & \textbf{0.88} & 0.76 & +0.12 \\ \hline
\texttt{train} & \textbf{0.81} & 0.73 & + 0.08 \\ \hline
\texttt{giraffe} & \textbf{0.98} & 0.83 & + 0.15 \\ \hline
\texttt{skis} & \textbf{0.62} & 0.74 & - 0.12 \\ \hline
\texttt{broccoli} &\textbf{0.88} & 0.58 & + 0.30\\ 

\end{tabular}
\caption{\textbf{Class Overall vs. CLIP Sample Accuracy.} The model's recall on the top-$100$ most similar images to prompt, ``\emph{a photo of [class]}'' using CLIP \citep{radford2021learning}'s cosine similarity, vs. all of the test set images that belong to each class.}
\label{table:class-sample-accs}
\end{table}

\paragraph{Version \#2: Compare the sample to the entire class}

In contrast to Version \#1, we also ran an ablation where we approximated the model's performance as the difference in accuracy on $\hat{\Phi}$, vs. the \emph{entire class} $D$ (\ie{} the class accuracies under the ``Overall'' column of Table \ref{table:class-sample-accs}).  As expected, because the model had \emph{lower} performance for four out of the five classes, this ablation resulted in a smaller proportion of hypotheses being deemed ``correct'' for a fixed threshold $\tau$.

\paragraph{Version \#3:  Compute contrastive similarity scores}
Our final ablation strategy modified Step \#1, the way we define which images are ``most similar'' to the hypothesis prompt.  
Specifically, rather than using the raw cosine similarity score between the individual hypothesis prompt and each image, we instead calculated the relative \emph{likelihood} that each image matched two candidate descriptions: (1) the hypothesis prompt, and (2) the class prompt.  We defined the ``most similar'' hypotheses as those that had the highest predicted probabilities of matching the hypothesis prompt. 

For example, if the users' hypothesis was ``\emph{a photo of a giraffe and zebra}'', then we would calculate the relative likelihood that each image ``matched'' the hypothesis prompt, vs. the class prompt (``\emph{a photo of a giraffe}'').

To calculate the final performance gap, we compare the model's performance on the matching $\hat{\Phi}$ to the entire class $D$ (like Version \#2).

We were motivated to try this variant for several reasons.  Intuitively, because we were already searching for photos that matched the hypothesis within a dataset of images that all belonged to the same class, including the class's name in the prompt wasn't really helping us narrow down on images that matched the hypothesis.
Because past work has shown that CLIP effectively functions as a bag-of-words model \citep{yuksekgonul2023visionlanguage}, searching with the hypothesis prompt alone may just return images of ``\emph{giraffes and zebras}'' that look extra giraffe-y (and don't even have the zebras that we want!).   

To examine the difference between the default vs. contrastive CLIP retrieval strategy, we ran a small-scale experiment where we retrieved the $80$ most similar images for $5$ hypotheses, $1$ per each class:
\begin{itemize}
    \item \texttt{a photo of water skis}
    \item \texttt{a photo of an airplane with people}
    \item \texttt{a photo of broccoli with other food}
    \item \texttt{a photo of a giraffe with zebras}
    \item \texttt{a photo of many people standing outside a train}
\end{itemize}

For each hypothesis, we examined a sample of the top-$80$ most similar images, and labeled all of the images that did or did not match.

We found that the model had a significant difference in accuracy for the two retrieval strategies: on average, the model had 35\% accuracy on the samples retrieved contrastively, vs. 65\% accuracy using the default strategy.  When we looked only at the examples that we labeled as \emph{matching} the hypothesis, the gap was more narrow but still existed: the model had 30\% accuracy on the contrastively retrieved images, vs. 47\% accuracy on the default strategy.

However, the contrastive retrieval strategy offered one major benefit: a much greater percentage of the images that were retrieved contrastively matched the hypothesis text description (75\% vs. 48\% for the default strategy).  
Retrieving contrastively to the class prompt was more likely to return matching images; but may result in an under-estimation of the model's true performance on all in-distribution images that match.

In conclusion, there are several reasonable strategies that have various pros and cons to retrieve a sample of images that matches a text description.
Specifically, while retrieving images contrastively to the class prompt does result in a higher-quality sample, the model has much lower accuracy on the retrieved sample.
Future work should continue to critically examine why and how the retrieval process used to find new images that match a hypothesis may mischaracterize the model's true behavior on the group.

\newpage
\section{Labeling Images that Match User Hypotheses}

\begin{figure}[h]
    \centering
    \includegraphics[width=0.7\linewidth]{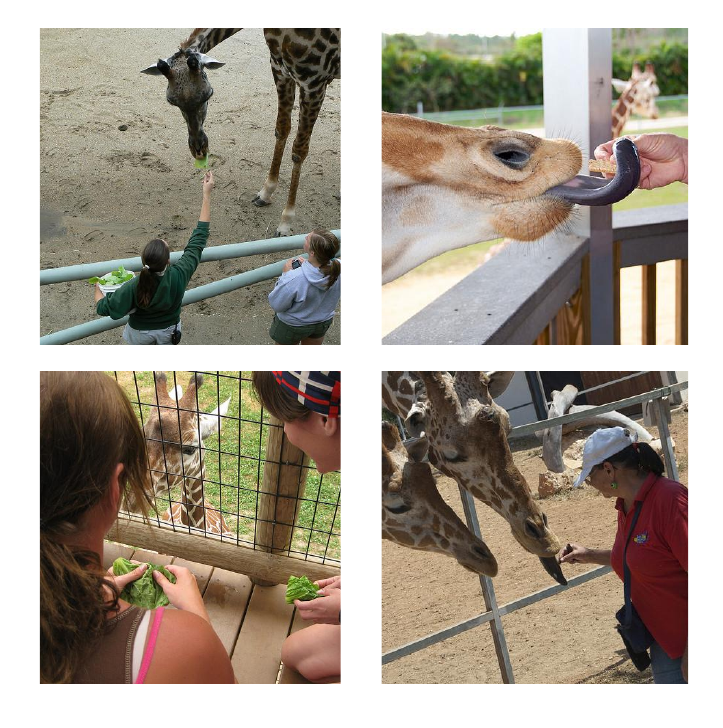}
\caption{Example images that a user selected as matching their hypothesis, ``\emph{a photo of people feeding a giraffe}'' }
\label{fig:example-matching-giraffe}
\end{figure}

To retrieve the sample $\hat{\Phi}$, we had to come up with a reproducible and consistent process to determine whether a new image matched the users' text description.
To do this, we created a \emph{labeling guide} inspired by \citep{shankar2020evaluating, vasudevan2022does} describing the criteria we used to determine whether an image matched each hypothesis.\footnote{We publicly release our labeling guide at \texttt{https://github.com/njohnson99/slice-discovery-human-eval}}

To ensure that our own labeling process was consistent with the users' intent, we checked that the criteria in our labeling guide were consistent with the user's labeling process by referencing the (up to $20$) images that the user selected as matching their description.
We only added labeling criteria that were consistent with the users' own labels.
We (the annotators) were blind to each group description's associated condition (algorithm) and the model's error rate on the selected images to avoid biasing our labeling.

We walk through an example to illustrate our labeling process.  Consider the hypothesis, ``\emph{a photo of people feeding a giraffe}''.  The user selected the images shown in Figure \ref{fig:example-matching-giraffe} as matching their hypothesis.  Given a sample of new images that all had giraffes, we had to determine which images belonged vs. did not belong to the users' group.  For this hypothesis, we came up with the following guidelines:

\begin{itemize}
    \item At least one person must be present in the image (\ie{} images of giraffes eating with no people in them do not belong to the group).
    \item Either (1) the person must be holding some food item (as in the bottom left image), or (2) the giraffe must be holding some food item in its mouth (as in the top left image) to belong to the group.  Images where people are standing near a giraffe, but there is no food, do not count.
\end{itemize}

Note that the above guidelines are consistent with the users' own labels of the images that matched their hypothesis.

\newpage
\section{Pilot Studies}

We ran several initial pilot studies to elicit feedback on our study instructions, interface, and design.
Our pilot studies demonstrated the importance of the instructions used when eliciting users' hypotheses.
Specifically, in an early pilot study, we did not prompt users to write down a single group corresponding to each slice, but instead allowed them to freely describe where they believed the model underperformed.
We found that when given this freedom, users often wrote down hypotheses that were the union of two distinct groups (\eg{} ``\emph{silhouetted giraffe OR giraffe with striped animal}''.  This behavior, while interesting, contradicted past works' assumption that each individual slice should only capture a single group where the model underperforms.
In the end, we decided to change our study instructions to elicit only a single group for each slice.

Our initial pilot study results informed our research hypotheses.  
Specifically, we were surprised to observe that the many users wrote down different hypotheses when we showed them the same slice in an early pilot study, where we showed the users only a single slice per each class.
This observation inspired our final research hypotheses.

\section{Training Details}

We fine-tune a pretrained (on ImageNet) ResNet-18 from the \texttt{torchvision.models} package.
We use the fine-tuning procedure proposed by \citet{kumar2022finetuning}, where we first linear probe (\ie{} hold all but the last linear layer as fixed) before we fine-tune (\ie{} optimize over all of the network weights).
We train using Adam with learning rate $0.001$.
We perform model selection by choosing the model with the best validation set cross-entropy loss.

\newpage
\section{Task Instructions}

\subsection{Task Description}

Below, we paste excerpts from the task description that was provided to users in the tutorial. \\

\texttt{Machine learning models can have blindspots, which occur when the model has lower accuracy on a coherent group of images (\ie{} the images in the group are united by a human-understandable concept).}

\texttt{Example. Given a dataset of images with tennis rackets, the model has much lower accuracy on images of "tennis rackets without people" (accuracy 41.2\%), compared to images of "tennis rackets with people" (accuracy 86.6\%).
Therefore, the model has blindspot "tennis rackets without people", because it has lower accuracy on images belonging to the group compared to images outside of the group.}

\texttt{In practice, after a model developer has trained a new object detection model, while the model may have blindspots, the model developer does not know what the model's blindspots are.}

\texttt{Why discover blindspots?:  There are several reasons why one may wish to discover an ML model's blindspots. For example,}

\begin{itemize}
    \item \texttt{Blindspots where the model underperforms on specific groups can contribute to algorithmic bias and cause downstream harm.}
    \item \texttt{Knowledge of where the model underperforms can inform decisions about deployment and actions model developers can take to fix (\eg{} re-train) the model.}
\end{itemize}

\texttt{Blindspot discovery methods are algorithms designed to help humans discover model blindspots. Given a dataset of images, the goal of a blindspot discovery method is to output "slices": groups of images that are both (1) high-error, and (2) coherent. Each slice is designed to correspond to a different true model blindspot.}

\texttt{However, the slices returned by these algorithms may not always be coherent. They may be noisy (\eg{} contain images that differ from the rest of the group), or fail to capture a coherent concept altogether.}

\texttt{In this study, you will be shown information about the model's performance on several slices output by a blindspot discovery algorithm.}

\texttt{For each slice, your primary goal is to describe the slice to the best of your ability by writing down a text description that:}
\begin{enumerate}
    \item \texttt{captures one distinct concept,}
    \item \texttt{that matches as many images in the slice as possible}
\end{enumerate}

\texttt{What does it mean to "describe" the slice?: }

\begin{itemize}
    \item \texttt{Your description should capture only one concept, even if it appears as though there are several different concepts represented in the slice.}
    \begin{itemize}
        \item \texttt{For example, if 5 of the images in the slice are of "tennis rackets with people" and 15 of the images in the slice are of "tennis rackets with dogs", you should write down "tennis rackets with dogs" (as this description matches more of the images in the slice).}
        
    \end{itemize}

    \item \texttt{The slice might be noisy or incoherent: it may be difficult to find a single slice description that matches all of the images in the slice. In these scenarios, you should try to write down a description that describes as many of the images as possible - preferably, at least 5 images (but the more the better!).}

    \item \texttt{Your slice description should be as clear, not subjective, and un-ambiguous as possible. Another person should be able to read what you write down, and quickly determine if a new image belongs to the group. }
\end{itemize}

\texttt{Some example slice descriptions for the tennis racket example are:}
\begin{itemize}
    \item \texttt{tennis rackets without people}
    \item \texttt{tennis rackets on clay courts}
    \item \texttt{tennis rackets without a tennis ball}
    \item \texttt{photos of tennis rackets taken inside of a residence}
\end{itemize}

\texttt{Examples of bad slice descriptions:}

\begin{itemize}
    \item \texttt{tennis rackets with dogs or tennis rackets with people.  Problem: This description is the "or" of two distinct concepts, when you are only supposed to have one!
Potential fix: "tennis rackets with dogs"} 
    \item \texttt{small tennis rackets. Problem: This description is ambiguous: what is a "small" vs. "big" tennis racket?
Potential fix: "tennis rackets designed for children"}
\end{itemize}

\texttt{Why describe each slice?}

\texttt{Because each slice was designed to correspond to a true model blindspot, being able to describe (in words) the true blindspot enables stakeholders to understand and communicate about actions they can take to address it.}

\texttt{Because blindspot discovery algorithms only return groups of points, they were designed with the goal that humans can look at their output and make sense of what coherent concepts are captured by each slice.
For example, discovering that "the model has accuracy 35\% on all images in the dataset of tennis rackets without people" is much more informative than, "the model has accuracy 30\% on this set of 20 images returned by a blindspot discovery algorithm".}

\subsection{Task Instructions}

\begin{figure*}[h]
\centering
\caption{A screenshot of the instructions shown to participants before they complete the slice questionnaires for a new object class.}
\label{fig:instructions-before-qa}
\includegraphics[width=\linewidth]{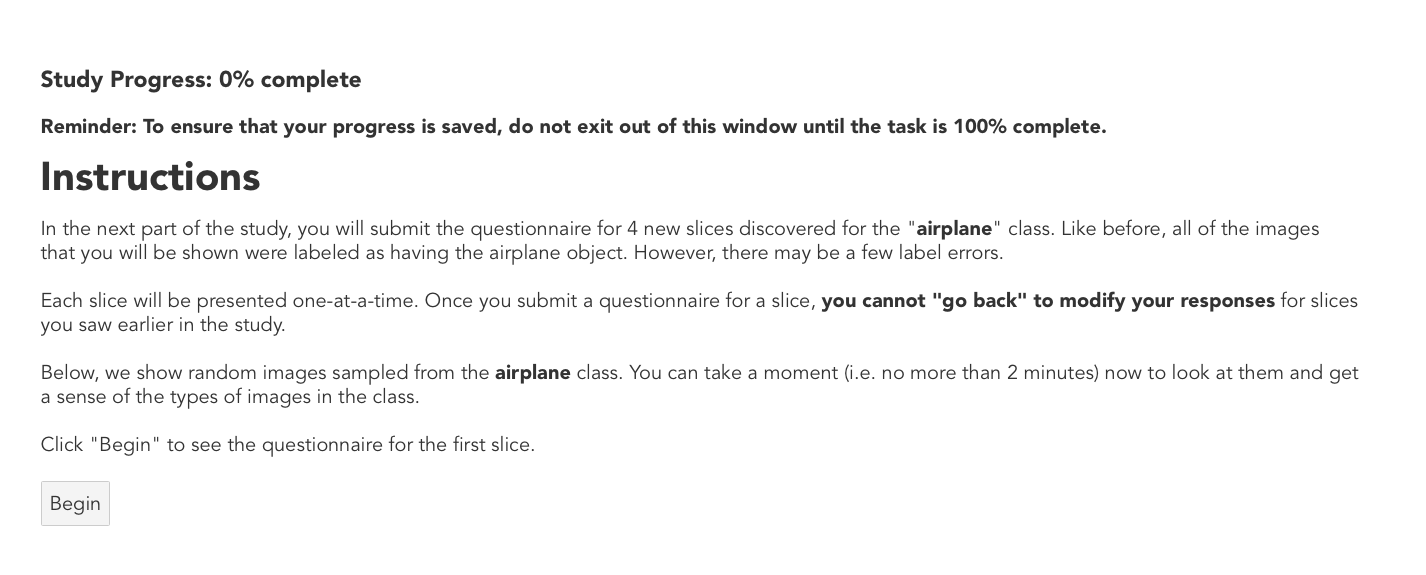}
\end{figure*}

See Figure \ref{fig:instructions-before-qa}.

\section{Slice Questionnaire}\label{apdx:slice-qa}

\begin{figure*}[h]
\centering
\caption{A screenshot of the slice questionnaire.}
\includegraphics[width=\linewidth]{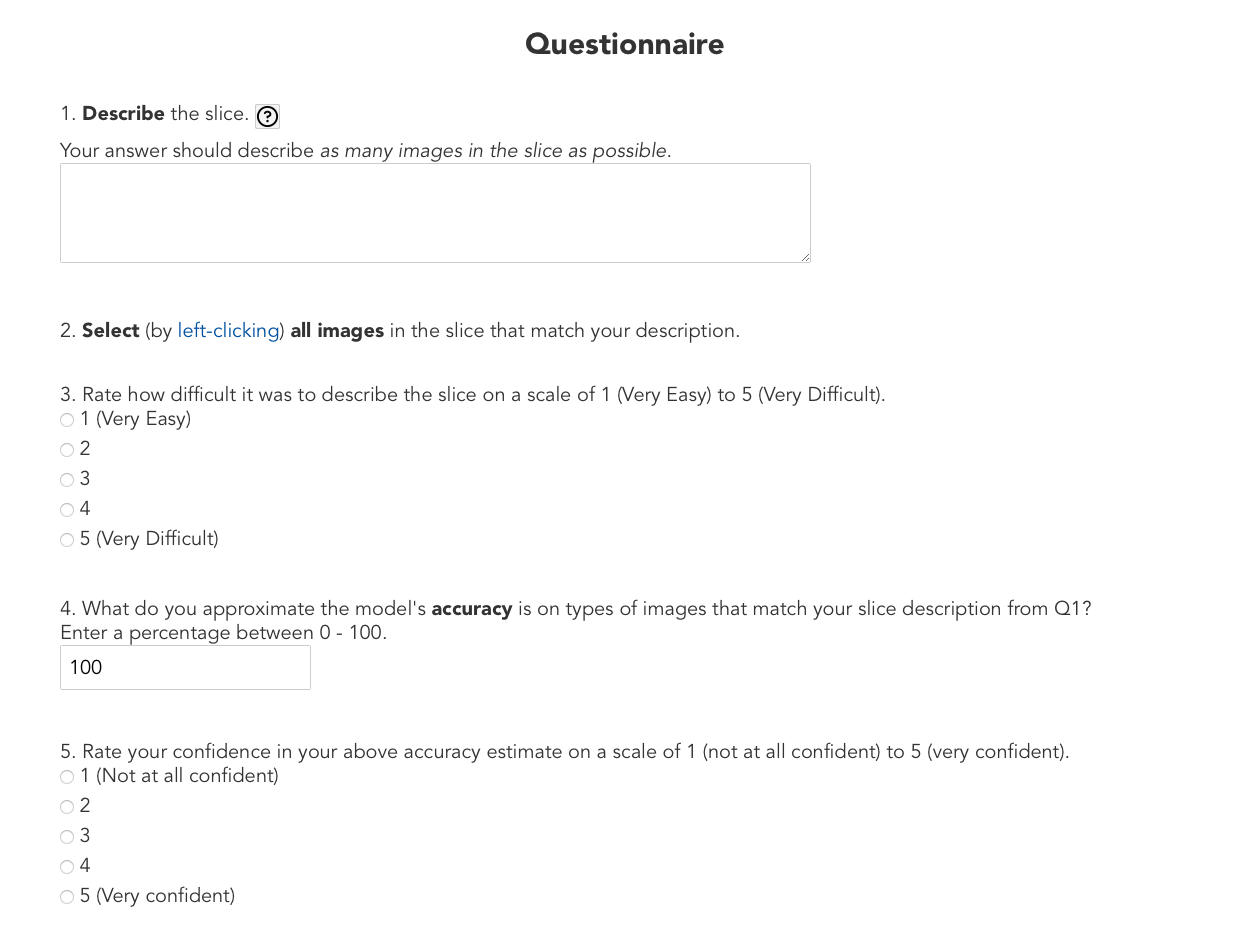}
\label{fig:slice-qa}
\end{figure*}

\begin{figure*}[h]
\centering
\caption{A screenshot of the tooltip text presented to users who click on the (?) icon next to Q1.}
\includegraphics[width=0.6\linewidth]{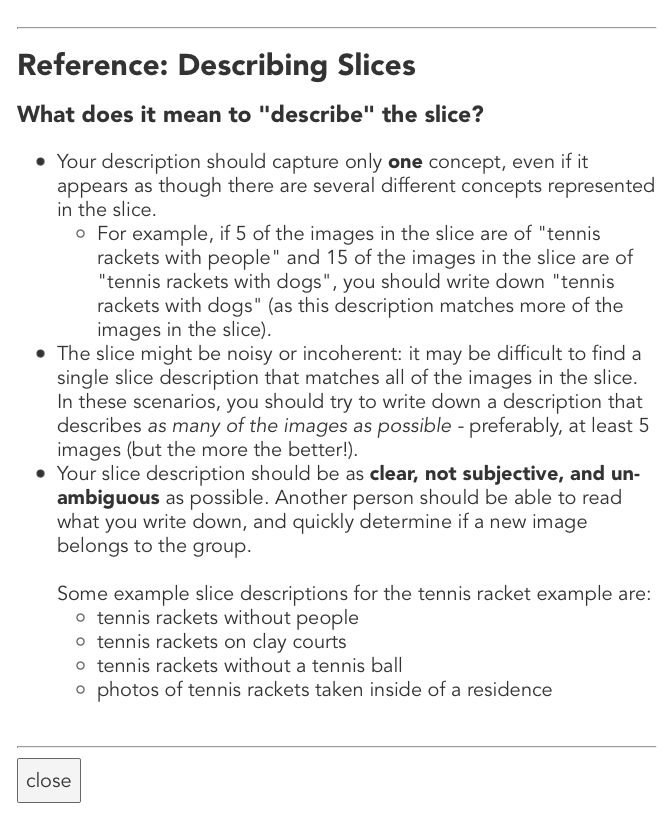}
\label{fig:tool-tip}
\end{figure*}

See Figures \ref{fig:slice-qa} and \ref{fig:tool-tip}.

\newpage
\section{Identifying Distinct Hypotheses: Extended}\label{apdx:user-variance}

For each slice, we asked two annotators to label whether pairs of hypotheses were \emph{synonymous}, \ie{} describe identical groups of points.  We presented the annotators with the following instructions: \\

\texttt{In this study, you will be shown lists of different group descriptions.  Each description was written by a human, and describes a group of images.}

\texttt{Your goal is to identify which descriptions are synonyms.  Synonymous descriptions may differ syntactically, but describe the same group of images.  If an image belongs to 1 group, it also will belong to all of its synonymous groups.}

\texttt{For example, the following two descriptions are synonyms, even though they differ syntactically:}

\begin{itemize}
    \item \texttt{D1:  "black-and-white photos of giraffes"}
    \item \texttt{D2:  "giraffes in greyscale"}
\end{itemize}

\texttt{The following two descriptions, while similar, are not synonyms, as there may be some images that would belong to D2 but not D1.}

\begin{itemize}
    \item \texttt{D1: "a photo of children eating broccoli"}
    \item \texttt{D2: "photos of broccoli and small children"}
\end{itemize}

Together, two annotators together annotated $48$ (out of $120$ possible) unique pairs of hypotheses as synonyms.
They disagreed (\ie{} only one annotator marked the pair as being a synonym) on $22$ of the pairs.
We define a pair of hypotheses as \emph{distinct} if \emph{neither annotator} noted that the pair was synonymous.  
In other words, if at least one annotator stated that the pair was synonyms, then we do not consider the hypotheses as being distinct.

\newpage
\section{Results: Extended}\label{apdx:extended-results}

Below, we detail the results of the statistical tests discussed in the main text, and present results from additional experiments when relevant.

\subsection{I.1 {}{} Correctness (Extended)}\label{apdx:correctness-ext}

Below we present the complete results of the statistical test for $\textbf{H1}$.

We ran an ANOVA test with Tukey post-hoc tests for multiple comparisons.  We compared the indicator for whether each hypothesis was correct with gap threshold $\tau = 0.2$ for each algorithm condition (\ie{} \textsc{Baseline}, \text{Domino}, \textsc{PlaneSpot}) corresponding to the hypothesis.  
We excluded hypotheses where we failed to find at least $15$ matching images (to approximate the performance gap) from our analysis. 
We retained $51, 44$, and $43$ rows corresponding to the \textsc{Baseline}, \textsc{PlaneSpot}, and \textsc{Domino} conditions respectively.
Figure \ref{fig:correctness} visualizes the mean and standard error of the correctness indicators. 

The value of the ANOVA $F$ statistic was $7.281$, and we found a statistically significant difference between the conditions $(p = 0.00099)$.
We report each pair-wise $p$-value in Table \ref{table:correctness-pairwise}.

\begin{table}[h!]
\centering
\begin{tabular}{ c | c }
Conditions & $p$-value \\ \hline
\textbf{\textsc{Domino} - \textsc{Baseline}} & $\mathbf{0.0007^*}$ \\
\textsc{PlaneSpot} - \textsc{Baseline} & $0.3756$ \\
\textsc{PlaneSpot} - \textsc{Domino} & $0.0594$ \\ 
\end{tabular}
\caption{\textbf{Hypothesis Correctness.} Results of Tukey post-hoc tests for pair-wise comparisons.  Significant $p$-values are denoted with an asterisk ($^*$).}
\label{table:correctness-pairwise}
\end{table}

\subsubsection{Ablation: Performance Gap Threshold}

\begin{figure*}[h]
\centering
\includegraphics[width=0.7\linewidth]{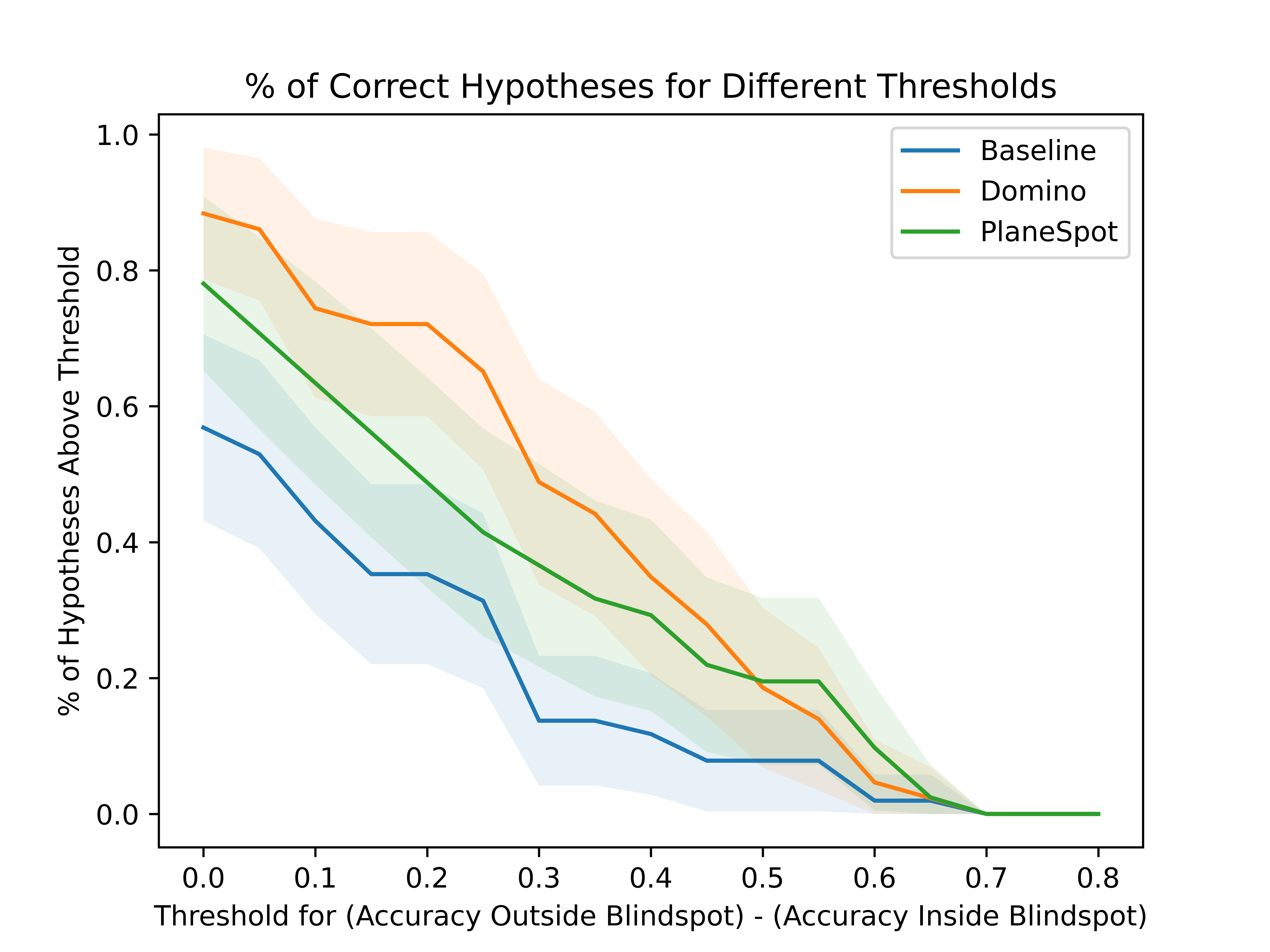}
  \caption{\textbf{Hypothesis Correctness.} The proportion of ``correct'' hypotheses per condition ($y$-axis), when we vary the performance gap threshold $\tau$ ($x$-axis).  The shaded region displays the 95\% confidence interval (defined as $\pm 1.96 \times $ the standard error) for each group.}
  \label{fig:ablate-performance-threshold}
\end{figure*}

We repeat the same hypothesis testing procedure where we ablate the performance gap threshold $\tau$ used to calculate the ``correctness'' indicators.  Figure \ref{fig:ablate-performance-threshold} displays how the average proportion of correct hypotheses changes as we increase the performance gap threshold $\tau$.  For all $\tau \leq 0.5$, we observe that a greater proportion of user hypotheses are ``correct'' for slices output by the \textsc{Domino}, then \textsc{PlaneSpot}, then \textsc{Baseline} algorithms.  Table \ref{fig:ablate-performance-threshold} shows the $p$-values of pairwise comparisons between conditions at different thresholds $\tau$.   We observe a statistically significant difference between the \textsc{Domino} and \textsc{Baseline} conditions for all thresholds $\tau \leq 0.4$.  There is a statistically significant difference between the \textsc{PlaneSpot} and \textsc{Baseline} conditions for $\tau = 0.3$ only.  There is no significant difference between the \textsc{PlaneSpot} and \textsc{Domino} conditions for any threshold.

\begin{table*}[h!]
\centering
\begin{tabular}{ c | c | c | c | c | c | c | c }
Conditions & $\tau = 0.1$ & $\tau = 0.15$ & $\tau = 0.2$ & $\tau = 0.25$ & $\tau = 0.3$ & $\tau = 0.35$ & $\tau = 0.4$   \\ \hline
\textbf{\textsc{Domino} - \textsc{Baseline}} & $\mathbf{0.0042^*}$ & $\mathbf{0.0006^*}$ & $\mathbf{0.0006^*}$ & $\mathbf{0.0019^*}$ & $\mathbf{0.0004^*}$ & $\mathbf{0.0018^*}$ & $\mathbf{0.0153^*}$ \\
\textbf{\textsc{PlaneSpot} - \textsc{Baseline}} & $0.1099$ & $0.0996$ & $0.3756$ & $0.5785$ & $\mathbf{0.0421^*}$ & $0.1301$ & $0.1247$ \\
\textsc{PlaneSpot} - \textsc{Domino} & $0.5444$ & $0.2489$ & $0.0594$ & $0.0538$& $0.3526$ & $0.3248$ & $0.7220$ \\ 
\end{tabular}
\caption{Results ($p$-values) of Tukey post-hoc tests for pair-wise comparisons.  Significant $p$-values are denoted with an asterisk ($^*$).}
\label{table:correctness-pairwise}
\end{table*}

\newpage
\subsection{I.2 {}{} Number of matching images (extended)}

Figure \ref{fig:matching-images} shows the mean, standard error, and distribution of the number of images that match each hypothesis.   The ANOVA $F$ statistic was $13.895$, and $p < 0.0001$.  We report each pair-wise $p$-value in Table \ref{table:num-matching-pw}.

\begin{table}[h!]
\centering
\begin{tabular}{ c | c }
Conditions & $p$-value \\ \hline
\textbf{\textsc{Domino} - \textsc{Baseline}} & $\mathbf{0.0000^*}$ \\
\textbf{\textsc{PlaneSpot} - \textsc{Baseline}} & $\mathbf{0.0000^*}$ \\
\textsc{PlaneSpot} - \textsc{Domino} & $0.9876$ \\ 
\end{tabular}
\caption{\textbf{Number of Matching Images.}  Results of Tukey post-hoc tests for pair-wise comparisons.  Significant $p$-values are denoted with an asterisk ($^*$).}
\label{table:num-matching-pw}
\end{table}

\newpage
\subsection{I.3 {}{} Self-reported difficulty (extended)}

Figure \ref{fig:reported-difficulty} shows the average self-reported difficulty of describing each slice, and a histogram of users' responses for each slice.  We report the $W$-statistic and $p$-value for each pair-wise comparison in Table \ref{table:difficulty-pw}. 

\begin{table}[h!]
\centering
\begin{tabular}{ c | c | c }
Conditions & $W$-statistic & $p$-value \\ \hline
\textbf{\textsc{Domino} - \textsc{Baseline}} & $W = 2798.5$ & $\mathbf{0.0000^*}$ \\
\textbf{\textsc{PlaneSpot} - \textsc{Baseline}} & $W = 2661.5$ & $\mathbf{0.0000^*}$ \\
\textsc{PlaneSpot} - \textsc{Domino} & $W = 1695$ & $0.5591$ \\ 
\end{tabular}
\caption{\textbf{Self-Reported Difficulty.}  Results of the pair-wise Mann-Whitey tests.  Significant $p$-values (accounting for the Bonferroni correction) are denoted with an asterisk ($^*$).}
\label{table:difficulty-pw}
\end{table}

\subsection{I.4 {}{} Correctness: CLIP Retrieval Ablation}

We also calculated the proportion of correct hypotheses for the two alternative approximation strategies detailed in Appendix B and ran our statistical tests using a correctness threshold of $\tau = 0.2$.  When we retrieved images \emph{contrastively}, we retained $160$ out of $180$ total hypotheses (89\%).  We found that for all retrieval strategies, a consistent trend holds: a larger proportion of \textsc{Domino} hypotheses are correct compared to \textsc{PlaneSpot}, which outperforms \textsc{Baseline} (Table \ref{table:correctness-ablation-props}).  
The difference between the \textsc{Domino} and \textsc{Baseline} conditions is significant for all thresholds.
The \textsc{Domino} condition significantly outperforms the \textsc{PlaneSpot} condition for the contrastive retrieval strategy only.

The percentage of all hypotheses that are correct for a fixed threshold $\tau$ varies across approximation strategies.  Approach \#2 appears to be more conservative in that only 37\% of all hypotheses are correct, which is significantly fewer than 68\% of all hypotheses for Approach \#3.  This difference aligns with our exploratory analyses in Appendix B, in which we observed that retrieving images contrastively tends to result in a less canonical (and thus more difficult) sample for a model to do well on.
In summary, while our finding of the relative ranking across conditions holds consistently for all approximation strategies, our results point to the difficulty of obtaining a representative sample $\hat{\Phi}$ using existing image retrieval tools.

Finally, our finding that slice coherence is uncorrelated with hypothesis correctness (\textbf{H4}) holds for all three approximation algorithms (Table \ref{table:ablation-h4}).

\begin{table}[h!]
\centering
\begin{tabular}{ c | c | c}
Approximation Strategy & $r_s$ & $p$ \\ \hline
Approach \#1 & $0.06$ & $0.5996$ \\ \hline
Approach \#2 & $0.08$ & $0.4728$ \\ \hline
Approach \#3 & $0.07$ & $0.5020$ \\ 
\end{tabular}
\caption{\textbf{Coherence vs. correctness for different performance gap approximation strategies}.  Reports the Spearman's rank correlation coefficient $r_s$ and $p$-value for the number of matching images (IV) vs. the indicator for whether the hypothesis is correct (DV), defining ``correctness'' using the three different approximation strategies detailed in Appendix B.}
\label{table:ablation-h4}
\end{table}

\begin{table*}[h!]
\centering
\begin{tabular}{ c | c | c | c}
Condition & Approach 1 & Approach 2 & Approach 3 \\ \hline
All & 0.51 (0.043) & 0.37 (0.04) & 0.68 (0.04) \\ \hline
\textsc{Baseline} & 0.35 (0.067) & 0.20 (0.06) & 0.48 (0.07) \\ \hline
\textsc{PlaneSpot} & 0.49 (0.079) & 0.41 (0.08) & 0.67 (0.07) \\ \hline
\textsc{Domino} & 0.73 (0.068) & 0.53 (0.08) & 0.89 (0.04) \\ 
\end{tabular}
\caption{\textbf{Correctness with CLIP Retrieval Ablations.} The percentage of hypotheses per condition that are ``correct'' (with standard errors) when we use samples $\hat{\Phi}$ retrieved by three different strategies detailed in Appendix B.  We use a performance gap threshold $\tau = 20\%$.  Percentages are calculated for the subset of hypotheses that we have a sufficiently large number of examples (\ie{} at least $15$ matching images) to approximate the performance gap.}
\label{table:correctness-ablation-props}
\end{table*}

\begin{table*}[h!]
\centering
\begin{tabular}{ c | c | c | c}
Conditions & Approach 1 & Approach 2 & Approach 3 \\ \hline
\textbf{\textsc{Domino} - \textsc{Baseline}} & $\mathbf{0.0007^*}$ & $\mathbf{0.00018^*}$ & $\mathbf{< 0.0001^*}$ \\
\textsc{PlaneSpot} - \textsc{Baseline} & $0.3756$ & 0.0690 & 0.0817 \\
\textsc{PlaneSpot} - \textsc{Domino} & $0.0594$ & 0.4661 & $\mathbf{0.0321^*}$ \\ 
\end{tabular}
\caption{\textbf{Correctness with CLIP Retrieval Ablations.} Results of Tukey post-hoc tests for pair-wise comparisons, comparing the proportion of correct hypotheses with threshold $\tau = 0.2$.  Significant $p$-values are denoted with an asterisk ($^*$).}
\label{table:correctness-ablation-tests}
\end{table*}

